\documentclass[12pt]{article}
\pdfoutput=1
\usepackage{jheppub}
\usepackage{amsmath}
\usepackage{bm}
\usepackage{slashed}
\usepackage{graphicx}
\usepackage{makecell}
\usepackage{enumitem}

\newlength{\dummysp}
\newcommand{\beq}{\begin{eqnarray}}
\newcommand{\eeq}{\end{eqnarray}}

\newcommand{\e}{{\epsilon}}
\newcommand{\s}{{\sigma}}

\newcommand{\gappeq}{\mathrel{\rlap {\raise.5ex\hbox{$>$}}
{\lower.5ex\hbox{$\sim$}}}}
\newcommand{\lappeq}{\mathrel{\rlap{\raise.5ex\hbox{$<$}}
{\lower.5ex\hbox{$\sim$}}}}

\newcommand{\ben}{\begin{enumerate}}
\newcommand{\een}{\end{enumerate}}

\newcommand{\bit}{\begin{itemize}}
\newcommand{\eit}{\end{itemize}}

\def\[{\left [}
\def\]{\right ]}
\def\({\left (}
\def\){\right )}

\def\Z{{\mathbb Z}}

\def\a{\alpha}
\def\b{\beta}

\def\e{\epsilon}

\def\L{\Lambda}
\def\m{\mu}
\def\n{\nu}

\def\r{\rho}
 
\def\s{\sigma}

\def\del{\partial}

\def\Z{\mathbb{Z}}
\def\lag{\mathcal{L}}

\def\ra{\rightarrow}

\def\bf{\textbf}
\def\it{\textit}
\def\f{\frac}

\def\a{\alpha}
\def\b{\beta}

\def\e{\epsilon}

\def\L{\Lambda}
\def\m{\mu}
\def\n{\nu}

\def\r{\rho}
 
\def\s{\sigma}

\def\del{\partial}

\def\Z{\mathbb{Z}}
\def\lag{\mathcal{L}}

\def\ra{\rightarrow}

\def\bf{\textbf}
\def\it{\textit}
\def\f{\frac}

\def\del{\partial}

\def\Z{\mathbb{Z}}
\def\lag{\mathcal{L}}

\def\mani{\mathcal{M}}

\title{Global aspects of $3$-form gauge theory: implications for axion-Yang-Mills systems}
\author{Mohamed M. Anber,}\author{Samson Y.L. Chan} 
  
\affiliation{Centre for Particle Theory, Department of Mathematical Sciences, Durham University, South Road, Durham DH1 3LE, UK}

\emailAdd{mohamed.anber@durham.ac.uk}\emailAdd{samson.y.chan@durham.ac.uk}    

\abstract{

{\flushleft{W}}e investigate the proposition that axion-Yang-Mills systems are characterized by a $3$-form gauge theory in the deep infrared regime. This hypothesis is rigorously examined by initially developing a systematic framework for analyzing $3$-form gauge theory coupled to an axion, specifically focusing on its global properties. The theory consists of a BF term deformed by marginal and irrelevant operators and describes a network of vacua separated by domain walls converging at the junction of an axion string. It encompasses $0$- and $3$-form spontaneously broken global symmetries. Utilizing this framework, in conjunction with effective field theory techniques and 't Hooft anomaly-matching conditions, we argue that the $3$-form gauge theory faithfully captures the infrared physics of the axion-Yang-Mills system. The ultraviolet theory is an $SU(N)$ Yang-Mills theory endowed with a massless Dirac fermion coupled to a complex scalar and is characterized by chiral and genuine $\mathbb{Z}_m^{(1)}$ $1$-form center symmetries, with a mixed anomaly between them. It features two scales: the vev of the complex scalar, $v$, and the strong-coupling scale, $\Lambda$, with $\Lambda \ll v$. Below $v$, the fermion decouples and a $U(1)^{(2)}$ $2$-form winding symmetry emerge, while the $1$-form symmetry is enhanced to $\mathbb Z_N^{(1)}$. As we flow below $\Lambda$, matching the mixed anomaly necessitates introducing a dynamical $3$-form gauge field of $U(1)^{(2)}$, which appears as the incarnation of a long-range tail of the color field. The infrared theory possesses spontaneously broken chiral and emergent $3$-form global symmetries.  It passes several checks, among which: it displays the expected restructuring in the hadronic sector upon transition between the vacua, and it is consistent under the gauging of the genuine $\mathbb Z_m^{(1)}\subset \mathbb Z_N^{(1)}$ symmetry.

 }

\begin{document}

\maketitle

%%%%%%%%%%%%%%%%%%
\section{Introduction}
%%%%%%%%%%%%%%%%%%%

Three-form gauge theory is a fascinating topic that attracted attention since L\"uscher's seminal work \cite{Luscher:1978rn}. There, it was argued that the non-trivial topology of the $4$-D Yang-Mills theory shows up in the infrared as an abelian long-range $3$-form gauge field $c_3$. It does not correspond to a physical massless particle, i.e., a propagating degree of freedom. Nevertheless, it contributes to the theory's vacuum energy, i.e., cosmological constant. In this formulation, the CP-violating $\theta$ term can be written as
\begin{eqnarray}
\theta \int_{{\cal M}_4} dc_3\,,
\end{eqnarray}
where $c_3$ is given in terms of the nonabelian Chern-Simons current density: $c_3=\mbox{tr}\left[\frac{1}{3} \left(a_1^c\right)^3+a_1^c \wedge da_1^c \right]/(8\pi^2)$, $a_1^c$ is the color field, and ${\cal M}_4$ is the $4$-D manifold. The correlator of the derivatives of two Chern-Simons
current densities is the topological susceptibility $\chi$ of Yang-Mills theory, which develops a pole as the momentum vanishes, the Kogut-Susskind pole \cite{Kogut:1974kt}, corresponding to a pole of the $c_3$ correlator. This is also known as the Veneziano ghost \cite{Veneziano:1979ec} because the pole appears with the opposite sign compared to those of conventional particles, and it provides an alternative means (to Yang-Mills instantons) to solve the axial $U(1)$ problem. One can write down an effective action that reproduces these findings \cite{DiVecchia:1980yfw}:
\begin{eqnarray}\label{Luscher action}
S_{\scriptsize \mbox{IR}}=\frac{1}{2\chi}\int_{{\cal M}_4} |dc_3|^2+\theta \int_{{\cal M}_4} dc_3\,,
\end{eqnarray}
where a kinetic-energy term of $c_3$ is added to give the correct long-range interaction form of the $c_3$ correlator. The effective action (\ref{Luscher action}) yields the vacuum energy \cite{Gabadadze:2002ff}:
\begin{eqnarray}\label{infinite N result}
E_0(\theta)=\frac{\chi}{2} \mbox{min}_k\left(\theta+2\pi k\right)^2\,,\quad k \in \mathbb Z\,.
\end{eqnarray}
This multi-branch function is periodic in $\theta$ and develops cusps at $\theta=\pi, 3\pi,...$, a result derived by Witten in the large-$N$ limit \cite{Witten:1980sp}. While this renowned result lends support to the validity of (\ref{Luscher action}), it is important to emphasize that the effective description (\ref{Luscher action}) is strictly derived in the large-$N$ limit.
See \cite{Gabadadze:2002ff} for a review and \cite{Aurilia:1980xj,Duff:1980qv,Freund:1980xh,Hata:1980hn, Nitta:2018yzb,Nitta:2018vyc} for further works on $3$-form gauge theory, including supersymmetric versions. 

When massless quarks are introduced, the $\theta$ term can be rotated away, and the theory restores its CP invariance, thus solving the strong CP problem. A similar effect can be achieved by introducing an axion $a$ through the Peccei–Quinn mechanism \cite{Peccei:1977ur}. This approach employs an anomalous global $U(1)$ symmetry along with a complex scalar field whose phase corresponds to the axion. The complex scalar undergoes spontaneous symmetry breaking at a scale $v$, which is much higher than the strong-coupling scale $\Lambda$. Below $\Lambda$, Yang-Mills instantons generate an effective potential for $a + \theta$, which is minimized at a point in the field space restoring the CP symmetry. The same conclusion can be reached using the $3$-form gauge theory, as demonstrated by Dvali in \cite{Dvali:2005an, Dvali:2022fdv}. In this framework, $c_3$ undergoes Higgsing when it absorbs the axion, resulting in $c_3$ becoming a short-range field. This mechanism can be elegantly seen in the Kalb-Ramond frame, where the axion is dualized to a $2$-form gauge theory. This process effectively eliminates the second term in (\ref{Luscher action}) and restores the CP invariance.

Since $c_3$ does not carry a physical degree of freedom, it is reasonable to question whether the $3$-form gauge theory is essential for formulating Yang-Mills theory in the deep IR (without or with axions) or if it is merely a redundant description lacking true physical significance. This work offers a new perspective on the validity of the $3$-form gauge theory beyond the large-$N$ limit. Under certain conditions, we argue that the $3$-form gauge theory is a faithful IR effective description in axion-Yang-Mills systems. This is achieved by examining the global symmetries of such systems as well as certain types of 't Hooft anomaly-matching conditions. 
 
 Our understanding of symmetries has undergone a conceptual paradigm shift over the past decade \cite{Gaiotto:2014kfa}; see \cite{Brennan:2023mmt} for a review and \cite{Hidaka:2019mfm,Brennan:2020ehu,Yokokura:2022alv,Yamamoto:2020vlk,Hidaka:2021kkf} for examples of works that discussed the $3$-form symmetries from a modern perspective. In the contemporary paradigm, a $p$-form global symmetry in $4$-D acts on a $p$-dimensional object and is generated by operators (symmetry defects) living on $(3-p)$-dimensional topological manifolds. Gauging a global symmetry is performed by introducing a background field of the symmetry and including an arbitrary sum over inequivalent classes of this background in the path integral. Moreover, it has been realized recently that the concept of symmetry can be extended to include operations that lack inversions; see, e.g., \cite{Kaidi:2021xfk,Cordova:2022ieu,Choi:2022jqy,Bhardwaj:2022lsg,Bartsch:2022mpm,Anber:2023pny} and the reviews \cite{Shao:2023gho,Schafer-Nameki:2023jdn}. Our primary goal is to conduct a systematic study of the $3$-form gauge theory and apply it to Yang-Mills theory coupled to matter fields and axions, with a focus on its global aspects. By utilizing newly developed mathematical tools, we can thoroughly examine the $3$-form effective field theory of axion-Yang-Mills systems. This effective description successfully passes several consistency checks.
 
 This work is divided into two main parts. The first part discusses the symmetry aspects of a general low-energy $3$-form gauge theory coupled to axion, including the multi-field case. Starting from a low-energy Lagrangian exhibiting a $U(1)^{(2)}$ $2$-form along with $U(1)^{(0)}$ $0$-form global symmetries, we construct a $3$-form gauge theory by gauging the former symmetry. The resulting theory is a topological quantum field theory (TQFT) of the BF type, modified by marginal and irrelevant operators, and describes a set of $q$ domain walls ($q$ is a free parameter) separating $q$ distinct vacua and forming a junction at the locus of an axion string. Generally, the gauge theory exhibits  $\mathbb{Z}_q^{(0)}$ $0$-form and $\mathbb{Z}_q^{(3)}$ $3$-form global symmetries, with a mixed anomaly between them valued in $\mathbb Z_q$. This anomaly is matched by breaking the two participating symmetries, leading to $q$ distinct vacua separated by domain walls. Furthermore, we point out that the theory encompasses a gauged $U(1)^{(-1)}$ $(-1)$-form symmetry, which undergoes spontaneous breaking (Higgsing), signifying that the vacuum energy has no contribution coming from $c_3$. We also construct the symmetry defects associated with these symmetries in two dual frames: the axion and the Kalb-Ramond frames. We demonstrate the action of such symmetries within an example of a domain-wall system. When the discrete $3$-form global symmetry or a subgroup thereof, $\mathbb Z_p^{(3)}\subseteq \mathbb Z_q^{(3)}$, is gauged,  we are left with only $q/p$ distinct vacua. 

In the second part of the paper, we apply this formalism to $SU(N)$ Yang-Mills theory endowed with a single massless Dirac fermion in a representation ${\cal R}$   coupled to a neutral complex scalar field. In the UV, this theory exhibits both a $\mathbb Z_{2T_{\cal R}}^{\chi(0)}$ $0$-form chiral and $\mathbb Z_{m}^{(1)}$ $1$-form center symmetries, with a possible mixed anomaly between the two symmetries. Here, $T_{\cal R}$ is the Dynkin index of ${\cal R}$ and $m=\mbox{gcd}(N,n)$, where $n$ is the $N$-ality of ${\cal R}$. When matched below the strong-coupling scale, this non-vanishing mixed anomaly necessitates introducing a dynamical $c_3$. We summarize the idea here, with details provided in the main body of the paper.
 
  Let $v$ and $\Lambda$ be the complex scalar vev and the Yang-Mills strong-coupling scales, respectively, and we take $\Lambda\ll v$. To see the $\mathbb Z_{2T_{\cal R}}^{\chi(0)}-\mathbb Z_{m}^{(1)}$ mixed anomaly, we turn on a $2$-form background field of $\mathbb Z_{m}^{(1)}$. This is achieved, as in \cite{Kapustin:2014gua,Gaiotto:2017yup,Seiberg:2018ntt,Anber:2020xfk},  by first introducing a pair of $1$-form and $2$-form $U(1)$ gauge fields $(B_1^c,B_2^c)$ along with the constraint $mB_2^c=dB_1^c$. The field strength of the $1$-form field satisfies the quantization condition $\int_{{\cal M}_2} dB_1^c\in 2\pi \mathbb Z$. Then the constraint implies that $B_2^c$ has a vanishing field strength $dB_2^c=0$, while its holonomy is fractional $\int_{{\cal M}_2} B_2^c\in \frac{2\pi \mathbb Z}{m}$. To couple the background field to fermions, in the second step, we enlarge $SU(N)$ to $U(N)$ and embed the $\mathbb Z_m^{(1)}$ background field into the $U(1)^{(1)}$ $1$-form symmetry of $U(N)$ gauge theory. However, we must ensure that the enlargement from $SU(N)$ to $U(N)$ does not introduce new degrees of freedom, which can be done by postulating that the theory is invariant under an auxiliary $1$-form gauge symmetry that acts simultaneously on the $U(N)$ and $(B_1^c,B_2^c)$ fields. As we flow to energy scales $\Lambda\ll E\ll v$, the magnitude of the complex scalar field freezes at $v$, while the winding of its phase (the axion) leads to the emergence of a $U(1)^{(2)}$ $2$-form symmetry that couples to the axion strings. Below $\Lambda$, the strong dynamics set in, leading to the confinement of the color field.  Here, one is faced with a puzzle: the confinement of the color field means that one should no longer incorporate the nonabelian field in the calculations. If true, the low-energy theory is no longer invariant under the postulated auxiliary $1$-form gauge symmetry, indicating that something is missing. The way out is to introduce the dynamical $3$-form gauge field $c_3$ of $U(1)^{(2)}$. The latter transforms non-trivially under the auxiliary gauge symmetry, ensuring the full low-energy effective field theory (in the background of the $\mathbb Z_{m}^{(1)}$ flux) is invariant under this auxiliary symmetry. Moreover, the IR theory reproduces the mixed anomaly, which is an important check on the analysis since anomalies are all-scale phenomena.  In this regard, $c_3$ can be thought of as the long-range tail of the nonabelian dynamics. Even though it does not carry a physical degree of freedom, its presence is essential for the consistency of the theory deep in the IR.  
    
  Below the scale, $v$, the fermions become massive, with a mass of order $v$, and decouple, leading to the enhancement of $\mathbb Z_m^{(1)}$ to $\mathbb Z_N^{(1)}$ $1$-form symmetry. The groups $U(1)^{(2)}$ and $\mathbb Z_N^{(1)}$ constitute a higher-group structure, where the former is the parent and the latter is the daughter symmetries. One may gauge the parent without gauging the daughter, but not conversely. As we flow below $\Lambda$, we may freely gauge $U(1)^{(2)}$  and introduce the dynamical $c_3$ without worrying about $\mathbb Z_N^{(1)}$. The latter stays an enhanced symmetry below $\Lambda$.

 One of our main results is the IR effective field theory at energy scale $E\ll \Lambda$ given by Eq. (\ref{low lag}), which we display here for convenience:
 \begin{eqnarray}\nonumber
{\cal L}_{E\ll\Lambda}=\frac{v^2}{2}da\wedge \star da+\frac{T_{{\cal R}}a}{2\pi}\left(dc_3-\frac{N}{4\pi}B_2^c\wedge B_2^c\right)+\Lambda^4{\cal K}\left(\frac{dc_3-\frac{N}{4\pi}B_2^c\wedge B_2^c}{\Lambda^4}\right)\,,\\
\end{eqnarray}
where ${\cal K}$ is the kinetic energy term of $c_3$, and the background of $\mathbb Z_m^{(1)}$ is activated. This theory exhibits $\mathbb Z_{T_{\cal R}}^{(0)}\times \mathbb Z_{T_{\cal R}}^{(3)}$ global symmetries. Dynamically, the IR theory forms axion domain walls, separating $T_{\cal R}$ distinct minima and breaking $\mathbb Z_{T_{\cal R}}^{(0)}$ and $\mathbb Z_{T_{\cal R}}^{(3)}$  maximally. The enhanced $\mathbb Z_N^{(1)}$ symmetry is explicitly broken by higher-order operators down to the genuine $\mathbb Z_m^{(1)}$ symmetry, which remains intact.

 The field strength of the $3$-form gauge field satisfies the quantization condition $\int_{{\cal M}_4} dc_3=2\pi m$, where $m$ is an integer equivalent to the topological charge of the Yang-Mills instantons.  The full partition function of the IR theory (at energy scale $E\ll \Lambda$) includes a sum over all integers $m$. We may integrate out $c_3$, and using the Poisson resummation formula,  we obtain the Euclidean partition function:
 \begin{eqnarray}\nonumber
Z[a]\sim\sum_{k\in \mathbb Z} \exp\left[-i\frac{k N}{4\pi}\int_{{\cal M}_4}B_2^c\wedge B_2^c\right]\exp\left[-\int_{{\cal M}_4}\frac{v^2}{2}da\wedge \star da+\frac{\Lambda^4} {8\pi^2}\left(T_{\cal R}a+2\pi k\right)^2\right]\,.\\
\end{eqnarray}
This partition function reproduces the chiral-center anomaly upon shifting $a\rightarrow a+\frac{2\pi}{T_{\cal R}}$. It also displays an infinite number of vacua, with the true vacuum energy given by
\begin{eqnarray}
V(a)\sim \Lambda^4\mbox{min}_k\left(T_{\cal R}a+2\pi k\right)^2\,.
\end{eqnarray}
The potential $V(a)$ has $T_{\cal R}$ minima at $2\pi \ell/T_{\cal R}$, $\ell=0,1,..., T_{\cal R}-1$, as well as cusps at $a=\pi(2\ell+1)/T_{\cal R}$, reflecting two facts. First, the cusps indicate that additional degrees of freedom, not accounted for by $V(a)$, are sandwiched between the true minima of the theory. These are the hadronic walls, which are very thin compared to the thickness of the axion domain walls \cite{Fugleberg:1998kk,Gabadadze:1999pp}. Second,  a restructuring in the hadronic sector occurs as one goes between one minimum and the other. These results are consistent with the large-$N$ limit (\ref{infinite N result}). 

 There exists a higher group structure between $\mathbb Z_{T_{\cal R}}^{(3)}$ and the enhanced $\mathbb Z_N^{(1)}$ symmetries. However, this structure trivialises for the genuine $\mathbb Z_m^{(1)}\subset \mathbb Z_N^{(1)}$ symmetry. This means we can gauge $\mathbb Z_m^{(1)}$ without worrying about $\mathbb Z_{T_{\cal R}}^{(3)}$. Gauging the former gives $SU(N)/\mathbb Z_m$ theory.  This theory still exhibits a spontaneously broken IR $\mathbb Z_{T_{\cal R}}^{(3)}$ symmetry due to the formation of domain walls. However, the chiral symmetry $\mathbb Z_{T_{\cal R}}^{(0)}$ becomes noninvertible. This results in dressing the domain walls with an IR TQFT. This intricate structure works as a consistency check on the adequacy of using the $3$-form gauge theory to describe the axion-Yang-Mills systems' IR physics.

This paper is organized as follows. In Section \ref{The Axion theory}, we set the stage by considering the field theory of a compact scalar, which possesses two global symmetries: shift and winding symmetries. The theory encounters a mixed 't Hooft anomaly, and thus, the shift symmetry breaks into a discrete group upon gauging the winding symmetry. Next, we couple the gauge field of the winding symmetry, the $3$-form gauge field, to the compact scalar and analyze the resulting theory in great detail: we identify the global symmetries, their mixed anomalies, and the noninvertible symmetries within this theory. Section \ref{The dual description} is devoted to studying the compact scalar in the dual frame, the Kalb-Ramond gauge theory, while Section \ref{Multi form gauge theory} generalizes these findings to two or more $3$-form gauge fields. In Section \ref{UV completion}, we use the machinery built in the previous sections to examine our proposal that the deep IR regime of the axion-Yang-Mills systems is described by a $3$-form gauge theory and apply various checks to this proposal. Finally, we conclude in Section \ref{Discussion} with a brief discussion. 

In this paper, dynamical fields (summed over in the path integral) are denoted by lowercase letters, while background fields use uppercase. We primarily use differential forms with subscripts indicating the form's degree.

%%%%%%%%%%%%%%%%%
\section{The Axion theory}
\label{The Axion theory}
%%%%%%%%%%%%%%%%%%

We consider the $4$-D theory of a $2 \pi$-periodic scalar field $a$, the axion, i.e., we identify $a({\cal P})\equiv a({\cal P})+2\pi \mathbb Z$ at the spacetime point ${\cal P}$. The basic Lagrangian is
\begin{equation}\label{axion Lag}
    \lag = \f{v^{2}}{2} da \wedge \star da\,,
\end{equation}
where $v$ is a constant of mass dimension $1$. When coupling the axion to a Yang-Mills theory via the Peccei-Quinn mechanism, $v$ is the axion's symmetry-breaking scale. The Lagrangian (\ref{axion Lag}) has a global $U(1)^{(0)}$ 0-form shift symmetry acting on the axion as $a\rightarrow a+\a$, where $\a$ is a constant. The corresponding Noether's $1$-form current is 
\begin{equation}
    j_{1} = v^{2} da\,,
\end{equation}
which is conserved thanks to the equation of motion:
\begin{equation}\label{0form current conservation}
    d \star j_{1}^{(0)} = v^{2} d \star d a = 0\,.
\end{equation}
The topological symmetry generator (symmetry defect) enacting this transformation is defined on a closed co-dimension-$1$ manifold $\mani_{3}$:
\begin{equation}\label{0form symmetry operator}
    U^{(0)}_{\a}(\mani_{3}) = e^{i \a \int_{\mani_{3}} v^{2} \star da}\,.
\end{equation}
The superscript emphasizes that the operator implements the action of a $0$-form symmetry. 
We can take $U^{(0)}_{\a}(\mani_{3})$ to surround  the local operator
\begin{eqnarray}V({\cal M}_0={\cal P})=e^{ia({\cal P})}\,,
\end{eqnarray}
 and then topologically deforming $ U^{(0)}_{\a}(\mani_{3})$ past $V$ to find  
\begin{equation}
    U^{(0)}_{\a} (\mani_{3}) V({\cal M}_0) = e^{i\a} V({\cal M}_0)\,.
\end{equation}

The axion theory is also endowed with a $U(1)^{(2)}$ 2-form global symmetry with a corresponding $3$-form current
\begin{equation}
    j_{3} = \star da\,,
\end{equation}
which is conserved because of the Bianchi identity:
\begin{equation}
    d \star j_{3} = d^{2} a = 0\,.
\end{equation}
We can also define the symmetry defect of the $U(1)^{(2)}$ $2$-form symmetry by integrating the Hodge-dual of $j_3$ on a co-dimension-$3$ manifold ${\cal M}_1$ as: 
\begin{eqnarray}
U^{(2)}_{\beta} (\mani_{1}) = e^{i\beta \int_{\mani_{1}} da}\,.
\end{eqnarray}
This symmetry defect acts on the $2$-dimensional axion-string worldsheet ${\cal M}_2$ \cite{Brennan:2020ehu}. Let $V(\mani_{2})$ be the axion-string Wilson surface, which has no local description\footnote{It is, however, possible to define an operator $e^{i\int_{\Sigma_3} v^2\star da}$ on an open surface $\Sigma_3$, with the string positioned at $\Sigma_2=\partial \Sigma_3$, the boundary of $\Sigma_3$. This approach mirrors the implicit definition of 't Hooft lines in earlier formulations. A direct definition of the Wilson surface operator as an integral over a closed $2$-dimensional surface will be given in Section \ref{The dual description} using the dual Kalb-Ramnod field.} in terms of the axion field $a$. 
Deforming the symmetry defect $U^{(2)}_{\b} (\mani_{1})$ past $V(\mani_{2})$  transforms the latter by a phase:
\begin{equation}
    U^{(2)}_{\b} (\mani_{1}) V(\mani_{2}) = e^{i  \beta \text{Link}(\mani_{1}, \mani_{2})} V(\mani_{2})\,,
\end{equation}
where $\text{Link}(\mani_{1}, \mani_{2})$ is the linking number between the two manifolds. 

There is a mixed 't Hooft anomaly between  $U(1)^{(0)}$ and $U(1)^{(2)}$ symmetries. To see it, we examine the commutation relation between the symmetry defects $U^{(0)}_{\a} (\mani_{3})$ and $U^{(2)}_{\beta} (\mani_{1})$. One way to perform the calculations is by foliating ${\cal M}_4$ into constant time slices and orienting both ${\cal M}_3$ and ${\cal M}_1$ to be time-like surfaces:
\begin{eqnarray}
U^{(0)}_{\a} (\mani_{3}(t))=e^{i\alpha v^2 \int_{\mani_{3}(t)} d^3x \partial_0 a(x,t)}\,,\quad  U^{(2)}_{\beta} (\mani_{1}(t))=e^{i\beta\int_{\mani_{1}(t)}\partial_i a dx^i}\,,
\end{eqnarray}
where $i\in\{1,2,3\}$. Then, using the equal-time commutation relation $\left[a(\bm x,t), \Pi_a(\bm y,t)\right]=i\delta^{(3)}(\bm x-\bm y)$, where $\Pi_a=v^2\partial_0 a$, and employing the Baker-Campbell-Hausdorff formula, we find
\begin{eqnarray}
U^{(0)}_{\a} (\mani_{3}(t))U^{(2)}_{\beta} (\mani_{1}(t))=e^{-i\alpha\beta}U^{(2)}_{\beta} (\mani_{1}(t))U^{(0)}_{\a} (\mani_{3}(t))\,.
\end{eqnarray}
The phase manifests the mixed anomaly: one cannot move the symmetry defects freely without encountering non-trivial phases.  The anomaly implies that gauging one symmetry breaks the other into, at most, a discrete subgroup.

%%%%%%%%%%%%%%%%%%%%%%%%%%%%%%%%%%%%%%%%%%%%%%%%%%%
\subsection{Gauging the $U(1)^{(2)}$ 2-form symmetry: $3$-form gauge theory and domain walls}
\label{Gauging the 2-form symmetry: Domain walls}
%%%%%%%%%%%%%%%%%%%%%%%%%%%%%%%%%%%%%%%%%%%%%%%%%%%

Now, we gauge the $U(1)^{(2)}$ symmetry, meaning that we introduce the $3$-form gauge field $c_3$ of the $2$-form symmetry and perform the path integral over $c_3$. We couple  $c_{3}$ to its current $j_3$ by adding the BF term 
\begin{eqnarray}\label{new term}
\f{q}{2\pi} \star j_{3} \wedge c_{3} = \f{q}{2\pi} da \wedge c_{3}
\end{eqnarray}
 to the Lagrangian (\ref{axion Lag}). We introduced the positive integer $q\in \mathbb N$ as a free parameter of the theory, and its physical significance will be apparent below.  The gauge field $c_3$ transforms as $c_3\rightarrow c_3+d\lambda_2$ under the $U(1)^{(2)}$ gauge transformation, and via integration by parts, we see that the new term (\ref{new term}) is invariant under this transformation. The field strength of $c_3$ is $f_4=dc_3$, and it satisfies the quantization condition:
 \begin{eqnarray}\label{quant f4}
 \int_{{\cal M}_4} f_4\in 2\pi \mathbb Z
 \end{eqnarray}
 on a closed ${\cal M}_4$. The consistency of the theory under $U(1)^{(2)}$ large gauge transformations implies that  $d\lambda_2$ satisfies the condition
 \begin{eqnarray}
  \int_{{\cal M}_3}d\lambda_2\in 2\pi \mathbb Z\,.
 \end{eqnarray}
 In addition, since $c_3$ is a dynamical field\footnote{The $3$-form gauge field has mass dimension $3$.}, we can include a kinetic energy term ${\cal K}$ for $c_3$. The total Lagrangian is\footnote{In a manifold with boundary, we also need to consider a boundary term so that the variation of the kinetic term at the boundary vanishes; see, e.g., \cite{Nitta:2018yzb} and references therein. We do not run into this subtlety in this work.}:
\begin{equation}\label{full axion lag}
    \lag = \f{v^{2}}{2} da \wedge \star da - \f{q}{2\pi} da \wedge c_{3} + \L^{4} {\cal K}\left(\f{dc_3}{\L^4}\right)\,,
\end{equation}
and we introduced the new scale $\Lambda$. More on ${\cal K}$  will be discussed momentarily. 
 The presence of the $c_{3}$ field reduces the $U(1)^{(0)}$ symmetry to a $\Z^{(0)}_{q} \subset U(1)^{(0)}$ symmetry; the Lagrangian (\ref{full axion lag}) is only invariant under the shift $a \ra a + 2\pi/q$. This demonstrates the earlier assertion that the mixed anomaly between the $U(1)^{(2)}$ and $U(1)^{(0)}$ symmetries leads to the latter being broken into a discrete subgroup when the former is gauged\footnote{As we shall see, this mixed anomaly is the IR incarnation of the axial-color ABJ anomaly in an axion-Yang-Mills UV complete theory.}. The current conservation law (\ref{0form current conservation}) of the $0$-form symmetry is modified to:
\begin{equation}
    v^{2} d \star da - \f{q}{2\pi} d c_{3} = 0\,.
\end{equation}
The corresponding $\mathbb Z_q^{(0)}$ symmetry defect is topological only when we include this combination of fields - (\ref{0form symmetry operator}) is modified to
\begin{equation}\label{sym of 0-form}
    U_{\a}^{(0)}(\mani_{3}) = e^{i \a \int_{\mani_{3}}\left( v^{2} \star da - \f{q}{2\pi} c_{3}\right)}\,.
\end{equation}
Using the quantization condition on $d\lambda_2$, i.e., $ \int_{{\cal M}_3}d\lambda_2\in 2\pi \mathbb Z$,  we readily see that $U^{(0)}_{\a}(\mani_{3}) $ is  gauge-invariant under a $U(1)^{(2)}$ gauge transformation if and only if $\a = 2\pi \ell/q, \ell\in \Z$. This seconds the above assertion that introducing the term (\ref{new term}) reduces the $U(1)^{(0)}$ symmetry down to a $\mathbb Z_q^{(0)}$ subgroup. 

Next, we focus on ${\cal K}$, the kinetic energy term of the $3$-form gauge field. In the following,  it will be helpful to write and analyze the Lagrangian (\ref{full axion lag}) in index notation\footnote{Translating from the $d$-forms to the index notation, it helps to remember that we are working in  Minkowski space, with metric $\eta_{\mu\nu}=\mbox{diag}(+1,-1,-1,-1)$, such that $\star\star=-1$ for even forms and $\star\star=+1$ for odd forms.}:
\begin{align}\label{full Lag in index notation}
    \lag & = \f{v^{2}}{2} \del_{\m} a \, \del^{\m} a -\f{q}{2\pi} \f{1}{3!}  \e^{\m\n\r\s}(\partial_\mu a)\, c_{\n\r\s} + \L^{4} {\cal K}\,,
\end{align}
where $\epsilon^{\mu\nu\alpha\beta}$ is the Levi-Civita tensor and the Greek indices run over $0,1,2,3$.
The canonical (quadratic) form of the kinetic energy term is:
\begin{eqnarray}
{\cal K}_{\scriptsize \mbox{can}}=-\frac{1}{2 \cdot4! \Lambda^8}f^{\mu\nu\alpha\beta}f_{\mu\nu\alpha\beta}\,,
\end{eqnarray}
where $f_{\mu\nu\alpha\beta}=\partial_\mu c_{\nu\alpha\beta}-\partial_\nu c_{\mu\alpha\beta}+\partial_{\alpha}c_{\mu\nu\beta}-\partial_\beta c_{\mu\nu\alpha}$. 
Since $f_{\mu\nu\alpha\beta}$ is totally anti-symmetric in the $4$ indices, we can always write it as
\begin{eqnarray}\label{form of f}
f_{\mu\nu\alpha\beta}=-\epsilon_{\mu\nu\alpha\beta}f(x)\,,
\end{eqnarray}
for some scalar function $f(x)$. Therfore, ${\cal K}_{\scriptsize \mbox{can}}$ takes the simple form\footnote{We used $\epsilon_{\mu\nu\alpha\beta}\epsilon^{\mu\nu\alpha\beta}=-4!$. We can also express $f$ in terms of $f_{\mu\nu\alpha\beta}$ as $f=\frac{1}{4!}\epsilon^{\mu\nu\alpha\beta}f_{\mu\nu\alpha\beta}$.}
\begin{eqnarray}
{\cal K}_{\scriptsize \mbox{can}}=\frac{f^2(x)}{2\Lambda^8}\,.
\end{eqnarray} 
The mathematical statement (\ref{form of f}) is equivalent to saying that the free $3$-form field $c_3$ does not carry propagating degrees of freedom. To see that, use the canonical kinetic term of $c_3$, ignore the axion in (\ref{full Lag in index notation}), and vary the Lagrangian with respect to $c_{\mu\nu\alpha}$ to find
\begin{eqnarray}
\partial_\mu f^{\mu\nu\alpha\beta}=0\,,
\end{eqnarray}
which admits the general solution $f_{\mu\nu\alpha\beta}=-\epsilon_{\mu\nu\alpha\beta}f$, where $f$, in this case, is a constant.  The constant field strength of a $3$-form gauge field carries no propagating degrees of freedom, much like free electrodynamics in $2$-D. In the absence of $a$,  the $4$-form field $f_4$ is a cosmological constant, which is easily seen by substituting $f_{\mu\nu\alpha\beta}=-\epsilon_{\mu\nu\alpha\beta}f$ into the Lagrangian\footnote{However, a subtle issue arises because the cosmological constant obtained in this manner does not align with the correct value derived by varying (\ref{full Lag in index notation}) with respect to the metric tensor \cite{Nitta:2018vyc}. This discrepancy can be resolved by including boundary terms in (\ref{full Lag in index notation}). Nonetheless, to obtain the correct form of the energy-momentum tensor and, consequently, the cosmological constant, we will rely on the variation of the action with respect to the metric tensor, disregarding boundary terms.}. 

From the perspective of effective field theory, the kinetic energy term can take a more generalized form, with ${\cal K}$ represented as a polynomial in $f$:
\begin{eqnarray}\label{generic KE}
{\cal K}\left(\frac{f}{\Lambda^4}\right)=\theta \frac{f}{\L^4}+\frac{f^2}{2\Lambda^8}+c'\frac{f^4}{\Lambda^{16}}+...\,,
\end{eqnarray}
where $\theta$ and $c'$ are some real parameters.The first term is topological, and since $\int_{{\cal M}^4}f_4=\int_{{\cal M}_4} d^4x f\in 2\pi \mathbb Z$, the theory is invariant under the shift $\theta\rightarrow \theta+2\pi$, and thus, $\theta$ is $2\pi$-periodic. This term breaks the CP invariance unless $\theta=\{0,\pi\}$. However, $\theta$ can be rotated away by combining $\theta \frac{f}{\L^4}$ with the second term in (\ref{full Lag in index notation}), after integrating by parts and shifting $qa\rightarrow qa-\theta$. Other possible higher-order kinetic energy terms may involve higher derivatives of $f_4$. Yet, these terms may be plagued with ghosts \cite{Nitta:2018vyc}, and thus, we ignore such terms in our construction.

By rescaling the fields $a$ and $f$ as $a \rightarrow av$ and $f \rightarrow \Lambda f$, we observe that the Lagrangian (\ref{full Lag in index notation}) manifests as a BF theory deformed by marginal terms (the canonical kinetic terms of $a$ and $f$) and by irrelevant operators (the higher-order terms of $f$). The full quantum theory guarantees that the system has $q$ degenerate ground states\footnote{We thank T. Sulejmanpasic and T. Tanizaki for discussions about this point.}. To see that, we may want to find an effective axion potential $V_{\text{eff}}(a)$ by integrating out $f_4$  and imposing the constraint (\ref{quant f4}). To streamline the analysis, we proceed by disregarding the axion kinetic energy term in (\ref{full axion lag}), which is a good approximation assuming $ \Lambda\ll v$. Then,  the Euclidean partition function reads:
\begin{eqnarray}\label{path int over c3}
Z[a]=\int [Dc_3]\sum_{m\in \mathbb Z}\delta\left(2\pi m-\int_{{\cal M}_4 } f_4\right)e^{-S_{E}}\,,
\end{eqnarray}
and 
\begin{eqnarray}
S_E=-\int_{{\cal M}_4}\Lambda^4{\cal K}\left(if_4/\Lambda^4\right)+i\frac{q}{2\pi}a\wedge f_4\,.
\end{eqnarray}
We can further simplify our analysis by recalling that $f_4$ does not carry propagating degrees of freedom and can be expressed by Eq. (\ref{form of f}). We take ${\cal M}_4$ to be a closed manifold, and therefore, we have $\int_{{\cal M}_4 } f_4=\int dV_{{\cal M}_4} f(x)=f_0V_{{\cal M}_4}$, where $f_0$ is the zero mode of $f(x)$ and $V_{{\cal M}_4}$ is the $4$-volume of ${\cal M}_4$, and we assume $\Lambda^4V_{{\cal M}_4}\gg 1$. Focusing only on the zero modes of $f$ and $a$, we find
\begin{eqnarray}\label{effective including sum m}
Z[a]\sim\left[\sum_{m\geq0}e^{\Lambda^4V_{{\cal M}_4}{\cal K}(2\pi i m)}\cos(mqa_0)\right]\times{\mbox{higher modes of}\,  a}\,,
\end{eqnarray}
where $a_0$ is the zero mode of $a$. 
 The higher modes of $a$ are suppressed by inverse powers of $\Lambda^4V_{{\cal M}_4}$ and can be neglected deep in the IR. The effective potential is defined via $V_{\text{eff}}(a)=-V_{{\cal M}_4}^{-1}\log Z[a]$, and thus, $V_{\text{eff}}(a_0)\sim \Lambda^4 {\cal F} (qa_0)$, where ${\cal F}$ is a periodic function with period $2\pi/q$.
We conclude that integrating out $f_4$ yields a periodic potential for the axion that respects the $\mathbb Z_q^{(0)}$ shift symmetry, as it should be. Minimizing $V_{\text{eff}}(a_0)$ yields $q$-degenerate ground states connected via domian walls. 

An alternative way to perform the path integral in (\ref{path int over c3}) is to use the Poisson resummation formula $\sum_{m\in \mathbb Z}\delta\left(2\pi m-\int_{{\cal M}_4} f_4\right)=\sum_{k\in \mathbb Z}e^{-ik\int_{{\cal M}_4}f_4}$. Taking ${\cal K}$ in the canonical form ${\cal K}_{\scriptsize \mbox{can}}=f_4^2/(2\Lambda^2)$, focusing on the zero modes,  and performing the Gaussian integral, we obtain
\begin{eqnarray}\label{alt der}
Z[a]\sim\sum_{k\in \mathbb Z}e^{-\frac{\Lambda^4 V_{{\cal M}_4}}{8\pi^2}\left(qa+2\pi k\right)^2}\,,
\end{eqnarray}
which, again, is a periodic function that respects the $\mathbb Z_q^{(0)}$ shift symmetry. The result (\ref{alt der}) is remarkable. This partition function displays an infinite number of vacua, most of which are false. The true vacuum energy is given by 
 \begin{eqnarray}
V(a)\sim \Lambda^4\mbox{min}_k\left(q a+2\pi k\right)^2\,.
\end{eqnarray}
This potential has $q$ distinct minima, with cusps at $a=\pi/q$, $3\pi/q$, etc. These cusps appear only upon including the infinite sum over $m$ in (\ref{effective including sum m}) and performing the Poisson resummation formula. In other words, the cusps are a feature of the full quantum theory. This observation will have far-reaching consequences in axion-Yang-Mills systems.

Let us return to the Lagrangian (\ref{full Lag in index notation}) and study its classical aspects.   Varying it with respect to $c_{\nu\rho\sigma}$,  assuming the general form of ${\cal K}$, yields the equation of motion of the $3$-form field:
\begin{equation}\label{3form eof}
    \f{q}{2\pi} \del_{\m} a =-\Lambda^4\del_{\m}{\cal K}'\left(\frac{f}{\Lambda^4}\right)\,,
\end{equation}
where the prime denotes the derivative with respect to the argument of ${\cal K}$. Equation (\ref{3form eof}) is pivotal to our subsequent analysis. Integrating once, we obtain
\begin{equation}\label{integrating once}
    \f{q}{2\pi} (a - \tilde a_{0}) =-\Lambda^4 {\cal K}'\left(\f{f}{\L^4}\right)\,,
\end{equation}
for some integration constant $\tilde a_{0}$. Assuming ${\cal K}'$ is invertible, we can rearrange the equation of $f$:
\begin{equation}\label{relation between for form and K}
    f = \L^{4} ({\cal K}')^{-1} \left(\f{q}{2\pi\Lambda^4} (\tilde a_0 - a) \right)\,.
\end{equation}
From  (\ref{full Lag in index notation}), the equation of motion for the axion $a$ is
\begin{equation}
    v^{2} \del_{\mu} \del^{\mu} a - \f{q}{2\pi} f = 0\,.
\end{equation}
Here, $f$ acts as the derivative of a classical effective potential for the axion field: in the presence of a classical effective potential for the axion, the equation of motion is $ v^{2} \del_{\mu} \del^{\mu} a+    \f{\del V_{\text{cl-eff}}(a)}{\del a}=0$. This means we can set $ \f{q}{2\pi} f =- \f{\del V_{\text{cl-eff}}(a)}{\del a}$, and using (\ref{relation between for form and K}), we conclude
\begin{equation}\label{equality between different things}
    \f{\del V_{\text{cl-eff}}(a)}{\del a} =- \f{q}{2\pi}f =- \f{q}{2\pi} \L^{4} ({\cal K}')^{-1} \left(\f{q}{2\pi\Lambda^4} (\tilde a_0 - a) \right)  \,.
\end{equation}
Then, one can integrate (\ref{equality between different things}) to obtain an expression of $V_{\text{cl-eff}}(a)$. Unlike the effective potential obtained from the full partition function, the classical effective potential does not need to yield $q$ degenerate ground states. The form of the classical effective potential depends on the kinetic energy term for $c_{3}$. To elucidate this point, we consider two examples of the resulting $V_{\text{cl-eff}}(a)$: the quadratic and the cosine potentials. The corresponding kinetic energy functions  are \cite{Dvali:2005an}:
\begin{eqnarray}\label{kinetic energy K 1}
   V_{\text{cl-eff}}^{\scriptsize\mbox{quadratic}}(a) &=& \f{\Lambda^4}{2} (a - \tilde a_{0})^{2}  \iff {\cal K}_{\scriptsize \mbox{can}}\left(\frac{f}{\Lambda^4}\right) =\frac{q^2f^2}{2\Lambda^8}\,,
\end{eqnarray}   
and
\begin{eqnarray}\label{kinetic energy K 2}\nonumber
   V_{\text{cl-eff}}^{\scriptsize \mbox{cos}}(a) &=& \L^{4} (1 - \cos\left(n(a-\tilde a_0)\right)\iff \\ 
    {\cal K}_{\scriptsize \mbox{cos}}\left(\frac{f}{\Lambda^4}\right) &=&-1+  \sqrt{1 - \left(\frac{qf}{2\pi n \Lambda^4}\right)^2} + \f{qf}{2\pi n \Lambda^4} \arcsin \left(\f{q f}{2\pi n\Lambda^4} \right)\,.
\end{eqnarray}
The integration constant in ${\cal K}$ is chosen such that $  {\cal K}\left(\frac{f}{\Lambda^4}=0\right)$=0.

 The kinetic energy term (\ref{kinetic energy K 2}) is designed to produce  $V_{\text{cl-eff}}^{\scriptsize \mbox{cos}}(a)$, and when $n\in q\mathbb N$, it exhibits multiple-of-$q$ minima. These minima are located at values of $a$ satisfying $ \f{\del V_{\text{cl-eff}}^{\scriptsize \mbox{cos}}(a)}{\del a}=0$, and from Eq.(\ref{equality between different things}) we see that $f$ vanishes there; the $3$-form gauge field $c_3$ is gapped at these minima.   Expanding ${\cal K}_{\scriptsize \mbox{cos}}\left(\frac{f}{\Lambda^4}\right)$  to the leading order in $f$ about one of the minima results, up to a proportionality constant, in the canonical kinetic energy term (\ref{kinetic energy K 1}). This, however, does not imply that the canonical kinetic energy term fails to produce $q$-degenerate ground states. As previously discussed, regardless of the form of ${\cal K}$, the full partition function of the Lagrangian (\ref{full axion lag}), incorporating the quantization condition (\ref{quant f4}),  will always lead to $q$-fold degeneracy in the deep IR regime. Nonetheless, ${\cal K}_{\scriptsize \mbox{cos}}\left(\frac{f}{\Lambda^4}\right)$  proves invaluable as it facilitates connections with textbook examples of axion domain walls. One can think of it as a UV completion of the canonical kinetic energy.

%%%%%%%%%%%%%
\bf{ Domin walls}. In the following, we proceed to discuss the classical domain wall solutions in the $U(1)^{(2)}$ gauge theory. We shall use the effective cosine potential in (\ref{kinetic energy K 2}) to carry out our analysis. However, this section's conclusions also hold for arbitrary potential, i.e., for arbitrary forms of ${\cal K}$. 

The cosine potential yields $n$ vacua $a_\ell=\frac{2\pi \ell}{n}$, where  $\ell=0,1,,...,n-1$. There are $n$  domain walls separating the adjacent vacua with a kink-like profile given by (here, we may set $\tilde a_0=0$):
\begin{eqnarray}\label{kink pofile}
a(z)=\frac{2\pi \ell}{n}+\frac{4}{n}\mbox{arctan}\left(e^{n m_a z}\right)\,,\quad -\infty<z<\infty\,,
\end{eqnarray}
and $m_a=\frac{\Lambda^2}{v}$ is related to the axion mass (the actual mass of the axion is obtained after using the canonical kinetic term, which yields the mass $\frac{n\Lambda^2}{v}$). We assumed the walls are space-filling in the $x$ and $y$ directions with a profile along the $z$-direction, taking ${\cal M}_4=\mathbb R^4$ for simplicity. We also assumed that the walls are separated by distances much larger than their width $\sim m_a^{-1}$. In the following, the statement $z\rightarrow \pm \infty$ means that $|z|\gg m_a^{-1}$ but still away from the adjacent walls. 
We observe that the reality of the kinetic energy term ${\cal K}$, see (\ref{kinetic energy K 2}), implies the inequality $|f|\leq 2\pi n\Lambda^4/q$. The value of ${\cal K}$ attains its minimum value, ${\cal K}=1$, at $f=0$, while it is maximized at $|f|=2\pi n\Lambda^4/q$. In addition, the first equality in (\ref{equality between different things}) yields:
\begin{eqnarray}
f(z)&=&-\frac{2\pi n}{q} \Lambda^4 \sin\left(na(z)\right)=-\frac{2\pi n}{q} \Lambda^4 \sin\left[4\mbox{arctan}\left(e^{n m_a z}\right)\right]\,.
\end{eqnarray}
The theory has a $\mathbb Z_{q}^{(0)}$ $0$-form symmetry that acts as $a\rightarrow a+\frac{2\pi}{q}$. The invariance of the cosine potential under the $0$-form symmetry demands that
\begin{eqnarray}
\frac{n}{q}\in \mathbb N\,,
\end{eqnarray}
and thus $n\geq q$. Later, we shall discuss that a stack of $n$ domain walls intersects at the locus of an axion string carrying a charge $q$ under $c_3$. It is more energetically favorable for the charge-$q$ string to support only $n=q$ domain walls. Nevertheless, we maintain the generality of $n$ and $q$ in our subsequent discussion\footnote{For example, we could have a kinetic energy term ${\cal K}$ that corresponds to a more general form of the effective potential $V_{\text{cl-eff}}(a)=\sum_{m\geq 1}\Lambda_m^4\left(1- \cos(mqa)\right)$.}.

\begin{figure}[h!]
  \centering
  \includegraphics[scale=0.5]{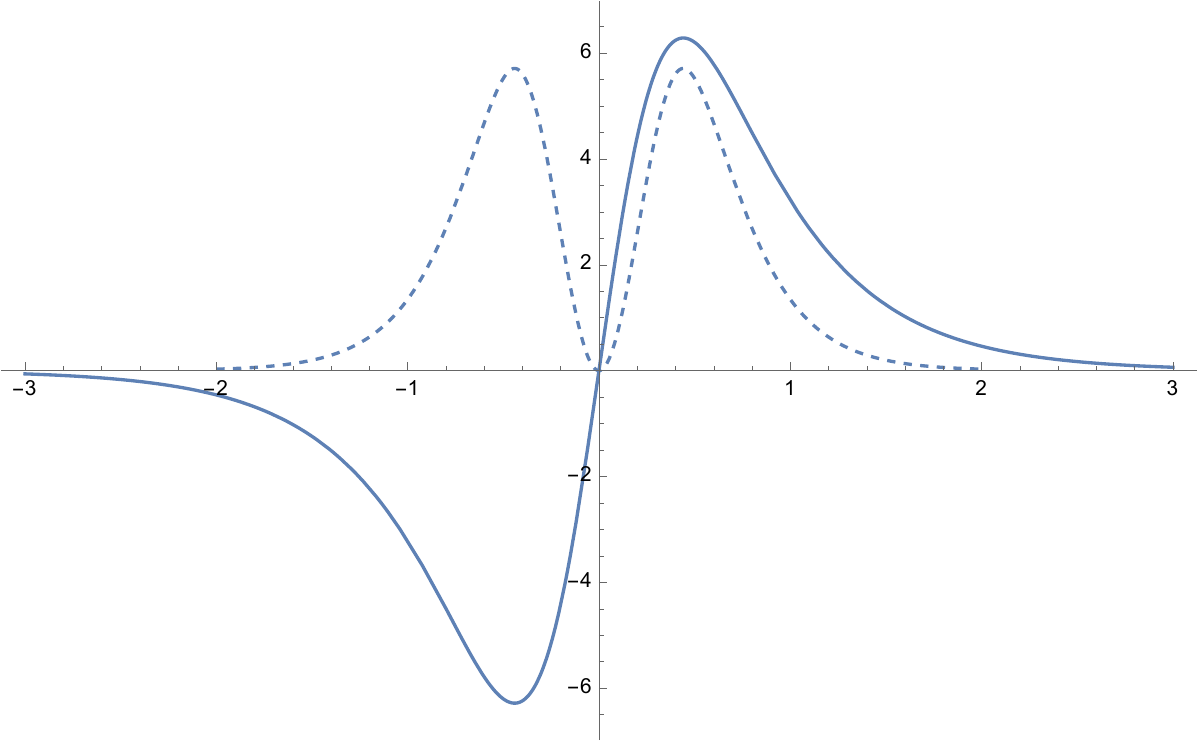}
  \caption{The profiles of $f$ (solid line) and   ${\cal K}_{\scriptsize \mbox{cos}}$ (dashed line) as functions of the coordinate $z$ (not to scale). We take $q=n=2$ and set $\Lambda=1$.}
  \label{Domain Walls graph}
\end{figure}

In the following, the derivatives of ${\cal K}$ will prove useful for our study:
\begin{eqnarray}
{\cal K}'=\frac{q}{2\pi n \Lambda^4}\mbox{arcsin}\left(\frac{qf}{2\pi n\Lambda^4}\right)\,, \quad {\cal K}''=\frac{q^2}{(2\pi n\Lambda^4)^2\sqrt{1-\left(\frac{qf}{2\pi n\Lambda^4}\right)^2}}\,.
\end{eqnarray}
Also, two important limiting behaviours of $f(z)$ are worth noticing:
\begin{eqnarray}\nonumber
f(z\rightarrow 0)&=&\frac{4\pi n^2 }{q}z  \Lambda^4\,,\\
f(z\rightarrow \pm \infty)&=& \frac{8\pi n}{q} \Lambda^4e^{-n m_a|z|}\mbox{sign}(z)\,.
\end{eqnarray}
At the wall core, $z=0$, we find that $f$ attains its mimimum value $f(0)=0$, where ${\cal K}$ is also minimized, while in the vacum $z\rightarrow \pm \infty$, we similarly have $f(\pm\infty)=0$, where again ${\cal K}$ is minimized. However, there exists a distance inside the wall, $\pm |z_{m}|$, at which $|f(z_m)|=\frac{2\pi n}{q}\Lambda^4$, i.e., it is maximized:
\begin{eqnarray}\label{def of zm}
m_a|z_m|=\frac{1}{n}\log\tan\left(\frac{\pi}{8}\right)\,.
\end{eqnarray}
At $ | z_m | $, the kinetic energy ${\cal K}$ is also maximized. Thus, $|f(z)|$ monotonically increases from the core of the domain wall until it reaches $z_m$, after which it starts decreasing exponentially; see Figure \ref{Domain Walls graph}. The exponential decay observed in $f(z)$ is a defining trait of a gapped system. In this scenario, the $4$-form field $f_4$ eats the axion, resulting in its acquisition of mass. This is also evident from 
\begin{eqnarray}\label{footnote for DW}
\int_{\mathbb R^4} f_4=0\,,
\end{eqnarray}
a result consistent with the quantization condition (\ref{quant f4}). In contrast,  in a gapless system, $|f(z)|$ would remain at the constant value of $\frac{2\pi n}{q}\Lambda^4$ from $\pm z_m$ to infinity. Below, we shall show that $\frac{q}{2\pi}\left(a(\infty)-a(-\infty)\right)=q/n$ is the domain wall charge under a $\mathbb Z_q^{(3)}$ $3$-form global symmetry. Thus, the stack of $n$ domain walls carries a total charge of $q$ under the $3$-form symmetry.

%%%%%%%%%%%%%%%%%%%%%%%%%%%%%%%%%%%%%%%%%%%%%%%%%
\subsection{The $3$- and $(-1)$-form symmetries, and their anomalies}
%%%%%%%%%%%%%%%%%%%%%%%%%%%%%%%%%%%%%%%%%%%%%%%%%

In this section, we show that the pure $U(1)^{(2)}$ gauge theory possesses a  $U(1)^{(3)}$ $3$-form global symmetry, which undergoes a breakdown into a $\mathbb Z^{(3)}_q$ symmetry in the presence of charge-$q$ matter. In addition, the theory is endowed with a gauged $(-1)$-form symmetry that can be used to diagnose the existence of a cosmological constant.  We further demonstrate that there is a mixed 't Hooft anomaly between $\mathbb Z^{(3)}_q$  and $\mathbb Z_q^{(0)}$ symmetries. Gauging the former gives rise to a $U(1)^{(2)}/\mathbb Z_{q}$ gauge theory.

%%%%%%%%%%%%%%%%%%%%%%
\bf{$3$-form global symmetry}. To begin, we rewrite Eq. (\ref{3form eof}), the equation of motion of $c_3$, in vacuum, i.e., setting the left-hand-side to $0$, as
\begin{eqnarray}
{\cal K}''\left(\frac{f}{\Lambda^2}\right)\partial_\mu f^{\mu\nu\alpha\beta}=0\,,
\end{eqnarray}
which implies either ${\cal K}''=0$ or $\partial_\mu f^{\mu\nu\alpha\beta}=0$. The first equation holds only when ${\cal K}$ is extremized, and thus, we concentrate solely on the latter equation that can be rewritten in the $d$-form language as 
\begin{eqnarray}
d\star dc_3=0\,.
\end{eqnarray}
This takes the form of the conservation law of the Hodge dual of a $4$-form current:
\begin{eqnarray}
d\star j_4=0\,, \quad \star j_4=\frac{\star dc_3}{\Lambda^4}=\frac{\star f_4}{\Lambda^4}\,,
\end{eqnarray}
implying that gauging the $U(1)^{(2)}$ $2$-form symmetry in vacuum gives rise to an emergent $U(1)^{(3)}$ $3$-form global symmetry. The $3$-form symmetry couples to $3$-surfaces ${\cal M}_3$, such that the Wilson surface operator is given by
\begin{eqnarray}\label{Wilson surface}
V({\cal M}_3)=e^{ip\int_{{\cal M}_3} c_3}\,,\quad p\in \mathbb Z\,.
\end{eqnarray}
The value $p=1$ gives the fundamental Wilson surface, while values of $p>1$ are higher representations. 
The conserved charge of this symmetry $Q^{(3)}$ is given by integrating $\star j_4$ over a $0$-dimensional manifold, or in other words, it is the local operator $\star j_4({\cal P})=\star f_4({\cal P})/\Lambda^4$ at the spacetime point ${\cal P}$. Using $\star f_4=\frac{1}{4!}\epsilon^{\mu\nu\alpha\beta}f_{\mu\nu\alpha\beta}$ along with $f_{\mu\nu\alpha\beta}=-\epsilon_{\mu\nu\alpha\beta} f$ and $\epsilon^{\mu\nu\alpha\beta}\epsilon_{\mu\nu\alpha\beta}=-4!$, we find
\begin{eqnarray}
Q^{(3)}=\star j_4({\cal P})=\frac{f({\cal P})}{\Lambda^4}\,,
\end{eqnarray}
and,  thus, the generator of the symmetry (symmetry defect) is
\begin{eqnarray}
U_{\gamma}^{(3)}({\cal M}_0={\cal P})=e^{i\gamma \star j_4}=e^{i\gamma f/\Lambda^4 }\,,
\end{eqnarray}
where $\gamma \in [0, 2\pi)$.
 The symmetry defect measures the amount of flux carried by a Wilson surface. Upon pushing $U_{\gamma}^{(3)}({\cal M}_0)$ past  $V({\cal M}_3)$, we obtain the algebra:
\begin{eqnarray}
U_{\gamma}^{(3)}({\cal M}_0)V({\cal M}_3)=e^{i p\gamma \text{Link}(\mani_{0}, \mani_{3}) } V({\cal M}_3)\,.
\end{eqnarray}

Let us repeat the analysis in the presence of matter. We shall use two approaches. First, we will investigate the $3$-form symmetry in the vicinity of the wall but far enough from its core. Next, we shall redo the analysis, this time without making any approximations or assumptions about the nature of the wall. We start with the approximate method, analyzing the situation near the domain walls of the cosine potential we studied above. We assume that we are far from the domain wall core, i.e., we are considering distances $|z|>z_m$, where $z_m$ is given by (\ref{def of zm}), which is the distance at which $f$ and ${\cal K}$ are maximized. At $z_m$, ${\cal K}''$ is ill-defined, invalidating our analysis; this is why we need to perform the calculations far from the core. Keeping this constraint in mind, we start by rewriting Eq. (\ref{3form eof}) in the $d$-form language as:
\begin{eqnarray}\label{d form of EOM of c3}
\Lambda^4 {\cal K}''(f) d\star dc_3=-\frac{q}{2\pi}da\,.
\end{eqnarray}
Far from the core, we can safely set $f\cong0$, as $f(z)$ decays exponentially fast at distances $z>z_m$. We have\footnote{Consider if we had employed the quadratic potential, as represented by Eq. (\ref{kinetic energy K 1}). In such a scenario, we would obtain ${\cal K}''=q^2/\Lambda^8$. The disparity of $(2\pi n)^2$ between these two cases becomes evident when we recognize that the function $f$ reaches its zero precisely at the local minimum of ${\cal K}$ in the case of the cosine potential, where we may approximate  ${\cal K}$ by a quadratic function ${\cal K}\cong 1+\frac{q^2f^2}{2(2\pi n)^2 \Lambda^8}$.}  ${\cal K}''(f\cong 0)=q^2/(4\pi^2n^2\Lambda^8)$. Using this information, we can rearrange (\ref{d form of EOM of c3}) as a conservation law:
\begin{eqnarray}\nonumber
d\star j_4=0\,, \quad Q^{(3)}=\star j_4({\cal P})&=&\frac{q^2}{(2\pi n)^2 \Lambda^2}\star f_4({\cal P})+\frac{q}{2\pi} a({\cal P})\\
&=&\frac{q^2}{(2\pi n)^2 \Lambda^2}f({\cal P})+\frac{q}{2\pi} a({\cal P})\,.
\end{eqnarray}
The symmetry defect is
\begin{eqnarray}\label{approximate U3}
U_{\gamma}^{(3)}({\cal M}_0={\cal P})=e^{i\gamma \star j_4}=e^{i\gamma\left(\frac{q^2 f (\cal P)}{(2\pi n)^2\Lambda^4}+\frac{q a({\cal P})}{2\pi}\right)}\,.
\end{eqnarray}
Since $a({\cal P})$ and $a({\cal P})+2\pi$ are identified, $U_{\gamma}^{(3)}({\cal M}_0={\cal P})$ is meaningful only when $\gamma \in 2\pi \mathbb Z/q\mathbb Z\equiv 2\pi \mathbb Z_q$. The charged object under the global symmetry is still the $3$-dimensional Wilson surface $V({\cal M}_3)$ given by (\ref{Wilson surface}). Now, the algebra of $U_{\ell}^{(3)}({\cal M}_0)$, $\ell \in \mathbb Z_q$, and  $V({\cal M}_3)$ is given by:
\begin{eqnarray}\label{the 3 form symmetry action}
U_{\ell}^{(3)}({\cal M}_0)V({\cal M}_3)=e^{i \frac{ 2\pi p\ell}{q} \text{Link}(\mani_{0}, \mani_{3}) } V({\cal M}_3)\,.
\end{eqnarray}

As we emphasized above, expression (\ref{approximate U3}) is valid only far from the domain wall core, i.e., (\ref{approximate U3}) is consistent with setting $f\cong 0$. However, inspired by the preceding analysis, we can repeat the treatment without making any approximations or assumptions about the nature of the walls. Our central equation, as usual, is (\ref{3form eof}), which we will write as a conservation law\footnote{In the cosine potential example, the derivative of $Q^{(3)}$ is ill-defined at the core. Nevertheless, $Q^{(3)}$ is well-defined everywhere.}:
\begin{eqnarray}\label{exact Q3}
\partial_\mu Q^{(3)}=0\,, \quad Q^{(3)}({\cal P})=\frac{q}{2\pi }a({\cal P})+\Lambda^4{\cal K}'\left(\frac{f({}\cal P)}{\Lambda^4}\right)\,,
\end{eqnarray}
with corresponding symmetry defect
\begin{eqnarray}\label{U of 3 form sym}
U_{\ell}^{(3)}({\cal M}_0)=e^{i\frac{ 2\pi \ell}{q}\left( \frac{q}{2\pi}a({\cal P})+\Lambda^4{\cal K}'\left(\frac{f({}\cal P)}{\Lambda^4}\right)\right)}\,, \quad \ell=1,2,..,q\,.
\end{eqnarray}
This is the generator of a $\mathbb Z_q^{(3)}$ symmetry for a generic form of the kinetic energy ${\cal K}$.
It is easy to check that (\ref{exact Q3}) reproduces the approximate expression (\ref{approximate U3}) of the cosine potential near $f\cong 0$. The charge $Q^{(3)}({\cal P})$ is a constant of motion unless one encounters a domain wall: crossing an elementary wall changes $Q^{(3)}$ by  
\begin{eqnarray}
\Delta Q^{(3)}=\frac{q}{2\pi}\left(a(\infty)-a(-\infty)\right)=\frac{q}{n}
\end{eqnarray}
units. In other words, $\Delta Q^{(3)}$ is the domain wall charge under the $\mathbb Z_q^{(3)}$ global symmetry. Interestingly, when $n=q$, the most natural scenario,  $\Delta Q^{(3)}$ coincides with the concept of topological charge in the theory of solitons. As we shall discuss in Section \ref{The dual description}, the domain walls intersect in a line; this is the locus of a string. From the flux conservation, this string carries a charge $q$, evenly distributed among the $n$ intersecting domain walls. We conclude that there are $n$ dynamical walls attached to a string, carrying a total charge $Q^{(3)}=q$ under $\mathbb Z_q^{(3)}$. 

The $3$-form symmetry $\mathbb Z_q^{(3)}$ acts on $q$ distinct Wilson surfaces $V({\cal M}_3)=e^{ip\int_{{\cal M}_3} c_3}$, $p=1,2,..,q$. When $p \neq 0$ Mod $q$, the flux carried by these Wilson surfaces cannot be absorbed by the dynamical domain walls since the latter always comes in a stack of a total charge $q$.

It remains to discuss the fate of the $3$-form global symmetry. We start with the $U(1)^{(3)}$ symmetry in the absence of matter, i.e., taking $v\rightarrow \infty$. In this case, as we discussed before, the equation of motion of the $3$-form gauge field $c_3$ yields a constant solution: $f_{\mu\nu\alpha\beta}=-\epsilon_{\mu\nu\alpha\beta}f$, where $f$ is a constant. Two walls experience a constant force, meaning that the  $U(1)^{(3)}$ is unbroken and the Wilson surface $V({\cal M}_3)$ exhibits the ``area" law $\langle V({\cal M}_3) \rangle=0$. Introducing the axion field and the coupling $-\frac{q}{2\pi}da\wedge c_3$ breaks $U(1)^{(3)}$ down to $\mathbb Z_q^{(3)}$ and endows the theory with a $\mathbb Z_q^{(0)}$ $0$-form global symmetry. Now, the $4$-form field $f_4$ is gapped, and the Wilson surfaces exhibit the ``perimeter" law $\langle V({\cal M}_3) \rangle\neq 0$, meaning that $\mathbb Z_q^{(3)}$ is spontaneously broken. Moreover, if the theory forms domain walls, $\mathbb Z_q^{(0)}$ also breaks spontaneously.

%%%%%%%%%%%%%%%%%%%%%%%%%%%%%%%%%%%%%%%%%%
\bf{The $(-1)$-form symmetry, its gauging, and the cosmological constant.} In addition, the theory possess a $U(1)^{(-1)}$ $(-1)$-form symmetry. The Bianchi identity $d^2c_3=0$ can be written as the conservation law of the Hodge-dual of a ``magnetic" current $\star j_4^m=dc_3=f_4$. The corresponding symmetry defect is
\begin{eqnarray}\label{-1 form sym}
U^{(-1)}_\gamma\left({\cal M}_4\right)=e^{i\gamma \int_{{\cal M}_4}f_4}\,,\quad \mbox{noting that}\quad \int_{{\cal M}_4}f_4 \in 2\pi \mathbb Z\,.
\end{eqnarray}
The operator $U^{(-1)}_\gamma\left({\cal M}_4\right)$ does not act on any physical objects directly. However, the two-point correlator $\langle c_3(x)c_3(0)\rangle$ can be used to determine whether the $(-1)$-form symmetry is preserved or spontaneously broken. A massless pole in the correlator signifies symmetry breaking; if absent, the symmetry remains unbroken \cite{Aloni:2024jpb}.

In the absence of axions, the $(-1)$-form symmetry functions as a global symmetry. The gauge field $c_3$ is massless, leading to a pole in its two-point correlator, which indicates that the $(-1)$-form symmetry is spontaneously broken. In this context, $c_3$ can be viewed as the Nambu-Goldstone field associated with this breaking.

When we couple $c_3$ to the axion through the term $\frac{q}{2\pi} a \wedge f_4$, the axion can be interpreted as the background gauge field for the $(-1)$-form symmetry. By introducing a kinetic term for $a$, we effectively sum over this background gauge field in the path integral, meaning we are gauging the $(-1)$-form symmetry\footnote{We thank an anonymous referee for pointing this out.}. As previously discussed, this leads to the axion acquiring mass, which can be understood as $a$ absorbing the would-be Goldstone field $c_3$ and becoming massive. Consequently, the gauged $(-1)$-form symmetry is spontaneously broken, and the correlator $\langle c_3(x)c_3(0)\rangle$ no longer exhibits a massless pole.

From our discussion, we find that the $(-1)$-form symmetry is intricately linked to the presence/absence of a cosmological constant sourced by $c_3$. The energy-momentum tensor of $c_3$ can be derived directly from (\ref{full axion lag}) by varying with respect to the metric tensor\footnote{Remember that we use the boundary condition  ${\cal K}(f=0)=0$.}:
\begin{eqnarray}
T_{\mu\nu}=\eta_{\mu\nu}\Lambda^4\left[{\cal K}\left(\frac{f}{\Lambda^4}\right)-\frac{f}{\Lambda^4}{\cal K}'\left(\frac{f}{\Lambda^4}\right) \right]\,,
\end{eqnarray}
which takes the form of a cosmological constant. Without axions, the global $(-1)$-form symmetry is spontaneously broken and $f=\mbox{constant}$, meaning that the vacuum energy gets a contribution from the long-range field $c_3$. Coupling to axions, the $(-1)$-form symmetry is gauged and Higgsed. Now, the system is gapped, i.e., $f=0$ (this is true far from the domain wall core); thus, the vacuum energy does not receive contribution from $c_3$, and we have $T_{\mu\nu}=0$.

%%%%%%%%%%%%%%%%%%%%%%%%%%%%%%%%%%%%%%%%%%%
\bf{The mixed 't Hooft anomaly between $\mathbb Z_q^{(3)}$ and $\mathbb Z_q^{(0)}$ symmetries.} An important question is whether there is a mixed anomaly between the two global symmetries $\mathbb Z_q^{(3)}$ and $\mathbb Z_q^{(0)}$. To answer this question, we examine the commutation relation between  $U_{\ell}^{(3)}({\cal M}_0)$ and $U_{\ell'}^{(0)}(\mani_{3})$ given by (\ref{U of 3 form sym}) and (\ref{sym of 0-form}), respectively. The calculations can be performed, as before, by foliating ${\cal M}_4$ into constant time slices and orienting ${\cal M}_3$  to be a time-like surface:
\begin{equation}
    U_{\ell'}^{(0)}(\mani_{3}(t)) = e^{i \frac{2\pi \ell'}{q} \int_{\mani_{3}(t)}\left( v^{2} \partial_0 a - \f{q}{2\pi} \frac{c_{ijk}\epsilon^{ijk}}{3!}\right)}\,.
\end{equation}
Then, using the equal-time commutation relation $\left[a(\bm x,t), \Pi_a(\bm y,t)\right]=i\delta^{(3)}(\bm x-\bm y)$, where $\Pi_a=v^2\partial_0 a- \f{q}{2\pi} \frac{c_{ijk}\epsilon^{ijk}}{3!}$, we obtain
\begin{eqnarray}\label{mixed anomaly 3 and 0 form}
    U_{\ell'}^{(0)}(\mani_{3}(t))  U_{\ell}^{(3)}({\cal M}_0(t))=e^{i\frac{2\pi \ell \ell'}{q}} U_{\ell}^{(3)}({\cal M}_0(t))  U_{\ell'}^{(0)}(\mani_{3}(t))  \,.
\end{eqnarray}
Thus, a $\mathbb Z_q$-valued mixed anomaly exists between the two symmetries. The anomaly implies one or both symmetries are broken. Our previous discussion reveals that both symmetries are broken in a theory that forms domain walls in the IR. 

We further explore the consequences of the mixed anomaly (\ref{mixed anomaly 3 and 0 form}), now working in a Hamiltonian formalism. Let $H$ be the Hamiltonian of the $U(1)^{(2)}$ gauge theory under study.  Since the theory has a $\mathbb Z_q^{(3)}$ $3$-form global symmetry, its generators commute with the Hamiltonian: $\left[H, U_{\ell}^{(3)} \right]=0$, and we can take the physical states in Hilbert space to be simultaneous eigenstates of these two operators. Thus, we have
\begin{eqnarray}\label{flux states}
|\psi\rangle_{\scriptsize\mbox{phy}}=|E(e), e\rangle\,, \quad e=0,1,.., q-1\,.
\end{eqnarray}
Here, $E(e)$ labels the energy and $e$ labels the ``flux" of the state (under the $3$-form symmetry)
such that 
\begin{eqnarray}\label{the eigenenergy and flux}
H|E(e), e\rangle=E(e)|E(e), e\rangle\,,\quad
 U_{\ell}^{(3)}|E(e), e\rangle=e^{i\frac{2\pi \ell e}{q}}|E(e), e\rangle\,.
 \end{eqnarray}
  Notice that the state's energy $E(e)$ can also depend on the value of the flux carried by the state. 
  
  To show that $e$ labels the flux carried by the state $|E(e), e\rangle$, let us insert a Wilson surface $V({\cal M}_3)=e^{ip\int_{{\cal M}_3} c_3}$ in the state $|E(e), e\rangle$ and then measure the new flux; the measurement is performed by acting with $U_{\ell=1}^{(3)}$ on the new state. Since this Wison surface carries a flux $p$, we expect inserting it increases the state flux by $p$ unit.  To confirm this, we perform the operation
\begin{eqnarray}\label{inserting a Wilson sur}
U_{\ell=1}^{(3)}V({\cal M}_3)|E(e), e\rangle=e^{i\frac{2\pi (e+p)}{q}}V({\cal M}_3)|E(e), e\rangle\,,
\end{eqnarray}
where we used (\ref{the 3 form symmetry action}) and (\ref{the eigenenergy and flux}). The relation (\ref{inserting a Wilson sur}) shows that the state $V({\cal M}_3)|E(e), e\rangle$ carries a flux $e+p$, and thus, indeed, $e$ in (\ref{flux states}) labels the flux carried by a state as anticipated. 

Next, we act with both sides of the anomaly (\ref{mixed anomaly 3 and 0 form}) on the state $|E(e), e\rangle$ to find that $U_{\ell'=1}^{(0)}|E(e), e\rangle$ is an eigenstate of $U_{\ell=1}^{(3)}$ with eigenvalue $e-1$. Since $\mathbb Z_q^{(0)}$ is a symmetry of the theory, we have $[H,U_{\ell'}^{(0)}]=0$, and thus, the state $U_{\ell'=1}^{(0)}|E(e), e\rangle$ has the same energy as $|E(e), e\rangle$. Repeating the statement $q$ times, we conclude that there are $q$ degenerate eigenstates of the same energy, labeled by the $q$ different values of $e$, as in (\ref{flux states}). This $q$-fold degeneracy is a direct consequence of the mixed anomaly (\ref{mixed anomaly 3 and 0 form}), which is true for both the ground and excited states of the system. In the thermodynamic limit, i.e., as we take the manifold ${\cal M}_4$ to be very large, we become interested mainly in the ground states, which stay $q$-fold degenerate. 

%%%%%%%%%%%%%%%%%%%%%%%%%%%%%%%%%%%%%%%%%%%%%%
\bf{Gauging $\mathbb Z_q^{(3)}$.} Next, we consider the $U(1)^{(2)}/\mathbb Z_{p}$ gauge theory resulting from gauging  a subgroup $\mathbb Z_p^{(3)}\subseteq \mathbb Z_q^{(3)}$. Gauging the discrete symmetry $\mathbb Z_p^{(3)}$ means introducing a background gauge field of the symmetry and including a sum over arbitrary insertions of this background in the path integral. We can introduce the  $\mathbb Z_p^{(3)}$ background $F_4$ into the path integral by replacing every $f_4$ in (\ref{full axion lag}) by $f_4\rightarrow f_4+F_4$, where $F_4$ can be expressed as $pF_4=dF_3$ and $dF_3$ satisfies the quantization condition $\int_{{\cal M}_4 } d F_3\in 2\pi \mathbb Z$. In turn, this implies the quantization of $F_4$ in units of $1/p$, i.e.,  $\int_{{\cal M}_4 } F_4\in \frac{2\pi}{p}\mathbb Z$, such that $dF_4$=0.

Now consider the second term in (\ref{full axion lag})  in the presence of the $\mathbb Z_q^{(3)}$ background. Its Euclidean form is  $i \f{q}{2\pi} a \wedge \left(f_4+F_4\right)$. Since, $\int_{{\cal M}_4 } F_4\in \frac{2\pi}{p}\mathbb Z$, only the shift $a\rightarrow a+ \frac{2\pi p}{q}$, i.e., $\mathbb Z^{(0)}_{q/p}$, survives as a genuine discrete symmetry. Actually, gauging $\mathbb Z_p^{(3)}$ renders the symmetry operator of $\mathbb Z_q^{(0)}$ a projective operator. To see this, consider the commutation relation between the $\mathbb Z_q^{(0)}$ symmetry defect $U_{\ell}^{(0)}(\mani_{3})$ and the generator of the $\mathbb Z_p^{(3)}$ symmetry $\left[U_{1}^{(3)}({\cal M}_0)\right]^{q/p}$ in the original $U(1)^{(2)}$ gauge theory. From (\ref{mixed anomaly 3 and 0 form}) we have
\begin{eqnarray}
\left[U_{1}^{(3)}(\mani_{0})\right]^{-q/p}  U_{\ell}^{(0)}({\cal M}_3) \left[U_{1}^{(3)}(\mani_{0})\right]^{q/p}  =e^{i\frac{2\pi\ell }{p}}  U_{\ell}^{(0)}({\cal M}_3)\,.
\end{eqnarray}
This relation shows that $U_{\ell}^{(0)}({\cal M}_3) $ fails to be a gauge-invariant operator in the $U(1)^{(2)}/\mathbb Z_p$ gauge theory. To remedy the problem, we sum over an arbitrary number of gauge transformations under the operator $\left[U_{1}^{(3)}(\mani_{0})\right]^{q/p}$ by defining the new $U(1)^{(2)}/\mathbb Z_p$ gauge-invariant symmetry defect:
\begin{eqnarray}\label{noninvert 0 form sym}\nonumber
 {\cal U}_{\ell}^{(0)}({\cal M}_3) &\equiv&\sum_{r\in \mathbb Z} \left[U_{1}^{(3)}(\mani_{0})\right]^{-rq/p}  U_{\ell}^{(0)}({\cal M}_3) \left[U_{1}^{(3)}(\mani_{0})\right]^{rq/p}\\\nonumber
&=&U_{\ell}^{(0)}({\cal M}_3) \sum_{r\in \mathbb Z} e^{i\frac{2\pi \ell r}{p}}\\
&=&U_{\ell}^{(0)}({\cal M}_3)\delta_{\ell\in p\mathbb Z}\,.
\end{eqnarray}
Thus, $ {\cal U}_{\ell}^{(0)}({\cal M}_3) $ is a projective operator.
 Recalling $\ell=0,1,..,q$, we conclude that the subgroup $\mathbb Z_{q/p}^{(0)}$ survives as a genuine symmetry\footnote{We thank E. Poppitz for discussion about this point.}. 

The $U(1)^{(2)}/\mathbb Z_p$ gauge theory has a remaining $\mathbb Z_{q/p}^{(3)}$ global symmetry with symmetry defect given by
\begin{eqnarray}\label{ remainingU of 3 form sym}
\left[U_{\ell}^{(3)}({\cal M}_0)\right]^{p}=e^{i\frac{ 2\pi \ell p}{q}\left( \frac{q}{2\pi}a({\cal P})+\Lambda^4{\cal K}'\left(\frac{f({}\cal P)}{\Lambda^4}\right)\right)}\,,\quad \ell=1,2,..,q/p.
\end{eqnarray}
The  $\mathbb Z_{q/p}^{(3)}$ symmetry acts on the $U(1)^{(2)}/\mathbb Z_p$ gauge-invariant Wilson surfaces $V({\cal M}_3)=e^{imp\int_{{\cal M}_3} c_3}$, $m\in \mathbb Z_{q/p}$. The action of $\mathbb Z_{q/p}^{(0)}$ takes us between  the $q/p$ vacua of the theory.

Let us discuss the consequences of gauging $\mathbb Z_p^{(3)}$ in the Hamiltonian formalism. When ${\cal U}_{\ell}^{(0)}({\cal M}_3)$ acts on the state $|E(e), e\rangle$, with $e=0,1,..,q-1$, it annihilates it unless $\ell$ is a multiple of $p$:
\begin{eqnarray}
 {\cal U}_{\ell}^{(0)}({\cal M}_3) |E (e), e\rangle=\delta_{\ell\in p\mathbb Z}U_{\ell}^{(0)}({\cal M}_3)|E(e), e\rangle\,,
\end{eqnarray}
 i.e., only $q/p$ states are not annihilated by ${\cal U}_{\ell}^{(0)}({\cal M}_3)$ . When we fully gauge $\mathbb Z_{q}^{(3)}$, i.e., in  $U(1)^{(2)}/\mathbb Z_q$ gauge theory, we have neither genuine $0$-form nor $3$-form  global symmetries. 

What has just happened, especially concerning a generic kinetic energy term ${\cal K}$, that leads to the formation of dynamical domain walls?  We analyze the situation by considering the cosine potential (\ref{kinetic energy K 2}), setting $n=q$ for simplicity. We shall also gauge the full $\mathbb Z_{q}^{(3)}$ symmetry. In this case, what we are operationally doing is that we are declaring the equivalence between all the $q$ vacua: $a\equiv a+ \frac{ 2\pi\ell }{q}$, $\ell=1,2,..,q$. Therefore, it is more meaningful to define $\varphi\equiv qa$ and replace $V_{\scriptsize\mbox{eff}}(a)=\Lambda^4\left(1-\cos(qa)\right)$ with $V_{\scriptsize\mbox{eff}}(\varphi)=\Lambda^4\left(1-\cos\varphi\right)$. The latter potential supports a single domain wall interpolating between $\varphi=0$ and $\varphi=2\pi$. However, such a wall is unstable quantum mechanically as it decays by instanton effects. We end up with a theory with a unique vacuum, supporting only strings but no domain walls.

%%%%%%%%%%%%%%%%%%%%%%%%%%%%%
\section{The dual description: the Kalb-Ramond frame}
\label{The dual description}
%%%%%%%%%%%%%%%%%%%%%%%%%%%%%%

While the axion framework provided valuable insights into the global symmetries within our system, the notion of strings remained implicit. To address this, we transition to the Kalb-Ramond frame \cite{Kalb:1974yc}, where the presence and properties of strings become more evident and accessible.

To dualize the axion Lagrangian (\ref{axion Lag}) to a theory of a $2$-form Kalb-Ramond field, we add to the Lagrangian  (\ref{axion Lag}) an extra term \cite{Dvali:2005an}:
\begin{equation}\label{axion with b}
    \lag = \f{v^{2}}{2} da \wedge \star da - \f{1}{2\pi} b_{2} \wedge d^{2} a\,.
\end{equation}
Here, $b_{2}$ is a Lagrange multiplier used to impose the Bianchi identity\footnote{The form of the extra term implies that $b_{2}$ has mass dimension $ 2$.}- integrating out $b_{2}$ gives $d^{2} a = 0$. We can also integrate out $a$ via its equation of motion:
\begin{equation}
    2\pi v^{2} \star da = db_{2}\,.
\end{equation}
Substituting back into (\ref{axion with b}) gives the dual Kalb-Ramond theory:
\begin{equation}
    \lag_{\text{dual}} = \f{1}{2 (2\pi)^{2} v^{2}} db_{2} \wedge \star db_{2}\,.
\end{equation}
The Kalb-Ramond field $b_{2}$ couples to the $2$-dimensional axion-string worldsheet $\mani_{2}$. Thus, the Wilson-like operator of an axion-string is: 
\begin{equation}\label{V of a string}
V({\mani_{2}}) = e^{i\int_{\mani_{2}} b_{2}}\,.
\end{equation}
Unlike $V({\mani_{2}})$, the operator $V({\cal M}_0)$, which in the original theory was given by $e^{ia}$, has no local description in terms of $b_2$. Therefore, in the Kalb-Ramond frame, $V({\mani_{2}})$ and $V({\cal M}_0)$ behave respectively like Wilson and 't Hooft operators in electrodynamics. This picture is reversed in the axion frame. 

In the dual description, the $0$-form and $2$-form global symmetry currents are:
\begin{equation}
    j_{1} = \f{1}{2\pi} \star db_{2}, \qquad j_{3} = 2\pi v^{2} db_{2}\,.
\end{equation}
These currents satisfy the conservation laws:
\begin{eqnarray}
d\star j_1=0\,,\quad d\star j_3=0\,,
\end{eqnarray}
which are the results of the Bianchi identity $d^2b_2=0$ and the equation of motion $d\star db_2=0$, respectively.
The corresponding $U(1)^{(0)}$ and $U(1)^{(2)}$ symmetry defects are given by
\begin{equation}\label{symmetries in KR frame}
    U^{(0)}_{\a}(\mani_{3}) = e^{i \a \int_{\mani_{3}} \f{1}{2\pi} db_{2}}\,,\quad     U^{(2)}_{\beta}(\mani_{1}) = e^{i \beta \int_{\mani_{1}} 2\pi v^{2} \star db_{2}}\,.
\end{equation}
With this dual formulation, we can see the action of $U(1)^{(2)}$ global symmetry- it shifts $b_{2}$ by a constant 2-form $\L_{2}$:
\begin{equation}
    b_{2} \ra b_{2} +  \L_{2}\,.
\end{equation}
 This transforms the axion string by a $U(1)$ phase:
\begin{equation}
    U_{\beta}^{(2)}(\mani_{2}) V({\mani_{2}})  = e^{i \beta \text{Link}(\mani_{1}, \mani_{2})} V({\mani_{2}}) \,.
\end{equation}
%

%%%%%%%%%%%%%%%%%%%%%%%%%%%%%%
\subsection{Gauging the $U(1)^{(2)}$ $2$-form symmetry}
%%%%%%%%%%%%%%%%%%%%%%%%%%%%%%

Here, we derive the Lagrangian of the dual Kalb-Ramond gauge theory, which results by gauging the $U(1)^{(2)}$ symmetry. Our starting point is the Lagrangian (\ref{full axion lag}) after adding the term $-b_2\wedge d^2a/(2\pi)$ to enforce the Bianchi identity $d^2a=0$:
\begin{equation}\label{full axion lag dual}
    \lag = \f{v^{2}}{2} da \wedge \star da - \f{q}{2\pi} da \wedge c_{3}-\frac{1}{2\pi}b_2\wedge d^2a + \L^{4} {\cal K}\left(\f{dc_3}{\L^4}\right)\,.
\end{equation}
The equation of motion of $a$ is:
\begin{eqnarray}
d \star da-\frac{1}{2\pi}d^2b_2-\frac{q}{2\pi}dc_3=0\,,
\end{eqnarray}
and integrating once we find
\begin{eqnarray}\label{inter a and b}
\star da=\frac{1}{2\pi v^2}\left(db_2+qc_3\right)\,.
\end{eqnarray}
Substituting (\ref{inter a and b}) into (\ref{full axion lag dual}), we obtain the dual Lagrangian
\begin{eqnarray}\label{KR manin Lag}
  \lag_{\text{dual}} = \f{1}{2(2\pi)^{2}v^{2}} (db_{2} + q c_{3})\wedge \star (db_{2} +q c_{3}) +\L^{4} {\cal K}\left(\f{dc_3}{\L^4}\right)\,.
\end{eqnarray}
In this formulation, the $U(1)^{(2)}$ symmetry is gauged by introducing the $3$-form gauge field $c_3$, which couples minimally to the Kalb-Ramond field $b_2$. The  minimal coupling $db_{2} + q c_{3}$ means that $b_2$ carries a charge $q$ under $c_3$. This is also manifest in the fact that the dual Lagrangian is invariant under the $U(1)^{(2)}$ local gauge transformation  $c_{3}\rightarrow c_3+d\lambda_2$, $b_2\rightarrow b_2-q\lambda_2$. 
 We may think of $b_2$ as the Stuckelberg field of $c_3$; as $c_3$ eats up the $b_2$ field, it aquires a mass $\sim\frac{\Lambda^2}{v}$. In the limit $v\rightarrow \infty$, the Kalb-Ramond field decouples, leaving us with a pure $U(1)^{(2)}$ gauge theory. 

  The effect of gauging the $U(1)^{(2)}$ symmetry is that the spectrum of extended operator changes. The Wlison surface operator (\ref{V of a string}) is no longer gauge-invariant under the $2$-form gauge symmetry. In the Kalb-Ramond frame, a gauge-invariant operator is
\begin{equation}
    e^{i \int_{\mani_{3}} db_{2} + q c_{3}} = e^{i \int_{{\cal M}_2=\del \mani_{3}} b_{2}} e^{iq\int_{\mani_{3}} c_{3}}\,,
\end{equation}
which can be interpreted as a string attached to a stack of domain walls with a cumulative charge of $q$ (the charge under the $c_3$ field).  An elementary axion that winds around this configuration cannot detect a nontrivial phase. This can be envisaged by computing the commutator
\begin{eqnarray}
\left[e^{ia},e^{i \int_{\mani_{3}} db_{2} + q c_{3}}\right]=0\,,
\end{eqnarray}
where we used (\ref{inter a and b}) along with $\left[a(\bm x,t), \Pi_a(\bm y,t)\right]=i\delta^{(3)}(\bm x-\bm y)$ and $\Pi_a=v^2\partial_0 a- \f{q}{2\pi} \frac{c_{ijk}\epsilon^{ijk}}{3!}$.

We end this section by discussing the global symmetries in the Kalb-Ramond frame. First, the generator of the $\mathbb Z_q^{(0)}$  $0$-form symmetry is given by $ U^{(0)}_{\a}(\mani_{3})$ in (\ref{symmetries in KR frame}). This generator, however, must be invariant under a $U(1)^{(2)}$ gauge transformation $b_2\rightarrow b_2-q\lambda_2$. Using $\int_{{\cal M}_3}d\lambda_2\in 2\pi\mathbb Z$, we find $\alpha=\frac{2\pi\ell}{q}$, $\ell=1,2,..,q$, as expected for the $\mathbb Z_q^{(0)}$ symmetry. Second, we also have a $\mathbb Z_{q}^{(3)}$ $3$-form global symmetry. However, the generator of this symmetry has no local description in the Kalb-Ramond frame.  It is crucial to highlight that transitioning between the axion and Kalb-Ramond frames does not eliminate global symmetries. Rather, certain symmetries may not be manifest in a local description. 

%%%%%%%%%%%%%%%%%%%%%%%
\section{Multi $3$-form gauge theory}
\label{Multi form gauge theory}
%%%%%%%%%%%%%%%%%%%%%%%

Consider an axion $a$ coupled to $2$ distinct $3$-form gauge fields $c_3$ and $\tilde c_3$. The Lagrangian reads
\begin{equation}\label{multi form lag}
    \lag =\frac{v^2}{2} |da|^2 - \f{q_{1}}{2\pi} da \wedge c_{3} - \f{q_{2}}{2\pi} da \wedge \tilde{c}_{3} + \Lambda^4{\cal K}\left(\frac{f_{4}}{\Lambda^4}\right) +\Lambda^4 \tilde{{\cal K}}\left(\frac{\tilde{f}_{4}}{\Lambda^4}\right)\,,
\end{equation}
where $f_{4}=dc_3$ and $\tilde{f}_{4}=d\tilde c_3$ are the field strengths of $c_{3}$ and $\tilde{c}_{3}$, respectively, and we assumed that the scale $\Lambda$ is the same for all the $3$-form fields. Notice that we did not include a kinetic-mixing term to avoid complications. The $3$-form gauge fields $c_3$ and $\tilde c_3$ are invariant under $U(1)^{(2)}\times U(1)^{(2)}$ gauge transformations $c_3\rightarrow c_3+d\lambda_2$ and $\tilde c_3\rightarrow \tilde c_3+d\tilde \lambda_2$, where $\int_{{\cal M}_3} d\lambda_2,\int_{{\cal M}_3} d\tilde\lambda_2\in 2\pi \mathbb Z$, while the field strengths satisfy the quantization conditions
\begin{eqnarray}
\int_{{\cal M}_4}f_4\,, \int_{{\cal M}_4}\tilde f_4\in 2\pi \mathbb Z\,.
\end{eqnarray}

The equations of motion of $a$, $c_3$, and $\tilde c_3$ read
\begin{eqnarray}\label{multi EOM}\nonumber
    &&v^{2} d \star da - \left(\f{q_1}{2\pi} d c_{3}+\f{q_2}{2\pi} d \tilde c_{3}\right) = 0\,,\quad    \f{q_1}{2\pi} \del_{\m} a =-\Lambda^4\del_{\m}{\cal K}'\left(\frac{f}{\Lambda^4}\right)\,,\\
  &&\f{q_2}{2\pi} \del_{\m} a =-\Lambda^4\del_{\m}\tilde {\cal K}'\left(\frac{\tilde f}{\Lambda^4}\right)\,.   
\end{eqnarray}
Trading $f$ and $\tilde f$ for a classical axion effective potential yields 
\begin{eqnarray}
  \f{\del V_{\text{cl-eff}}(a)}{\del a} =- \f{1}{2\pi}\left(q_1f+q_2\tilde f\right)\,. 
\end{eqnarray}
This relationship asserts that the combination of fields $q_1f+q_2\tilde f$ vanishes at the extrema of $V_{\text{cl-eff}}(a)$.  This implies that this particular combination of the $3$-form fields is gapped at the theory's vacua. Conversely, the independent combination $q_2f-q_1\tilde f$ remains ungapped \cite{Dvali:2005an}. 

 The system (\ref{multi form lag}) enjoys a multitude of global symmetries. First, the theory is invariant under a $\mathbb Z_q^{(0)}$ symmetry that acts on $a$ as $a\rightarrow a+\frac{2\pi}{q}$ and $q=\mbox{gcd}(q_1,q_2)$. The generator of $\mathbb Z_q^{(0)}$ is

\begin{equation}\label{sym of 0-form multi}
    U_{\ell}^{(0)}(\mani_{3}) = e^{i \frac{2\pi \ell}{q} \int_{\mani_{3}}\left( v^{2} \star da -\f{q_1 c_3+q_2\tilde c_3}{2\pi}\right)}\,,\quad \ell=1,2,...,q\,,
\end{equation}
which acts on the local operator $e^{ia({\cal P})}$.

In addition, the system exhibits two independent $3$-form global symmetries. To find them, we use two methods. We start with the first method, which was not discussed previously but works as an alternative view on the global $3$-form symmetry. In this method,  we shift $c_{3}$ and $\tilde{c}_{3}$ by two independent closed but not exact 3-forms $\Lambda_3$ and $\tilde\Lambda_3$:
\begin{equation}
    c_{3} \ra c_{3} + \L_{3}\,, \quad \tilde{c}_{3} \ra \tilde{c}_{3} + \tilde{\L}_{3}\,,
\end{equation}
under which the action gets shifted by
\begin{equation}
    S \ra S + \f{q_{1}}{2\pi} \int_{{\cal M}_4} da \wedge \L_{3} + \f{q_{2}}{2\pi} \int_{{\cal M}_4} da \wedge \tilde{\L}_{3} = S + q_{1} k \a + q_{2} k \b\,,
\end{equation}
where we defined $\int_{{\cal M}_3} \L_{3} = \a, \int_{{\cal M}_3} \tilde \L_{3} = \b$, and we recalled that $\int_{{\cal M}_1} da = 2\pi k, k \in \Z$ since $a$ is a compact scalar. Under the $\Lambda_3$ and $\tilde\Lambda_3$ shifts, the path integral picks up a phase:
\begin{equation}
Z \ra e^{iq_{1} k \a + iq_{2} k \b}Z\,.
\end{equation}
It is easily seen that there are two combinations of $\alpha$ and $\beta$ that lead to two independent $3$-form global symmetries that leave the action invariant: a $U(1)^{(3)}$ symmetry is obtained by setting $q_1\a = -q_2\b$ and a $\mathbb Z_{q}^{(3)}$ symmetry is obtained upon taking $\a, \b \in \f{2\pi}{q} \Z$, where $q = \gcd(q_{1}, q_{2})$. These are linearly independent transformations, so there are no redundancies, and the faithful 3-form symmetry group is 
\begin{equation}
     \Z_{q}^{(3)} \times U(1)^{(3)}\,.
\end{equation}

Another way to obtain the same result is by combining the equations of motion of $c_3$ and $\tilde c_3$ in the form of two independent conservation laws. Using (\ref{multi EOM}) we find
\begin{eqnarray}\nonumber
\partial_\mu\left(\frac{q_1a+q_2 a}{2\pi}+\Lambda^4{\cal K}'\left(\frac{f}{\Lambda}\right)+\Lambda^4\tilde {\cal K}'\left(\frac{\tilde f}{\Lambda}\right)\right)&=&0\,,\\
\partial_\mu\left(-q_2\Lambda^4{\cal K}'\left(\frac{f}{\Lambda}\right)+q_1\Lambda^4\tilde {\cal K}'\left(\frac{\tilde f}{\Lambda}\right)\right)&=&0\,,
\end{eqnarray}
from which we define the two symmetry defects:
\begin{eqnarray}\nonumber
U_{\alpha_1}^{(3)}\left({\cal M}_0\right)&=&e^{i\alpha_1\left(\frac{q_1a+q_2 a}{2\pi}+\Lambda^4{\cal K}'\left(\frac{f}{\Lambda}\right)+\Lambda^4\tilde {\cal K}'\left(\frac{\tilde f}{\Lambda}\right)\right)}\,,\quad
U_{\alpha_2}^{(3)}\left({\cal M}_0\right)=e^{i\alpha_2\left(-q_2\Lambda^4{\cal K}'\left(\frac{f}{\Lambda}\right)+q_1\Lambda^4\tilde {\cal K}'\left(\frac{\tilde f}{\Lambda}\right)\right)}\,.\\
\end{eqnarray}
While the phase $\alpha_2$ is an arbitrary $U(1)$ phase, implying that $U_{\alpha_2}^{(3)}$  is the symmetry defec of a $U(1)^{(3)}$ $3$-form global symmetry, the single-valuedness of $U_{\alpha_1}^{(3)}$ as $a\sim a+2\pi$ implies that $\alpha_1=\frac{2\pi \mathbb Z}{q}$, reducing the second symmetry group from $U(1)^{(3)}$ down  to $\mathbb Z_q^{(3)}$, in accordance with our earlier finding.

The theory also possesses two distinct $U(1)^{(-1)}$ $(-1)$-form symmetries associated with the Bianchi's identities: $d^2c_3=d^2\tilde c_3=0$. As we mentioned above, only the field combination $q_1c_3+q_2\tilde c_3$ is gapped while the other combination $-q_2c_3+q_1\tilde c_3$ remains gappless. This implies that only one of the two $(-1)$-form symmetries is gauged and spontaneously broken (Higgsed), resulting in the axion acquiring a mass. The other $(-1)$-form symmetry is a global symmetry, which also exhibits spontaneous breaking, resulting in a massless $3$-form gauge field that sources a cosmological constant in the deep IR. 

This treatment is easily generalized to any $K$ distinct $3$-form fields to find that the full faithful symmetry group is 
\begin{equation}
 \mathbb Z_q^{(0)}\times \Z_{q}^{(3)} \times \prod_{i=1}^{K-1} U(1)^{(i)(3)}
\end{equation}
where $q = \gcd(q_{1}, \ldots, q_{K})$.

%%%%%%%%%%%%%%%%%%%%%%%%%%
\section{UV completion: axion-Yang-Mills theory}
\label{UV completion}
%%%%%%%%%%%%%%%%%%%%%%%%%%%

In this section, we argue that the $3$-form gauge theory in either the axion or the Kalb-Ramond frame emerges in the IR from a UV-complete axion-Yang-Mills system\footnote{The reader might object referring to the system we study in this section as UV complete since we use a scalar field that exhibits a Landau pole. Here, by a UV-complete, we just mean a model that couples the axion to Yang-Mills theory and gives the desired symmetries and natural hierarchy of scales.}. This conclusion is reached by using effective field theory methods empowered by new 't Hooft anomaly matching conditions. We put the IR effective field theory under scrutiny by testing its adequacy under various checks. 

To this end, consider an $SU(N)$ gauge theory endowed with a massless Dirac fermion in a representation ${\cal R}$ under $SU(N)$. In addition, consider a complex scalar $\Phi$ that is inert under $SU(N)$ but otherwise couples to the Dirac fermion. The Lagrangian of the system reads \cite{Anber:2020xfk}:
\begin{eqnarray}\label{YMA lag}
{\cal L}=-\frac{1}{2g^2}\mbox{tr}\left(f_2^c\wedge \star f_2^c\right)+\bar\psi\bar\sigma^\mu D_\mu \psi+\bar{\tilde\psi}\bar\sigma^\mu D_\mu {\tilde\psi}+|d\Phi|^2-V(\Phi)+y\Phi\tilde \psi\psi+\mbox{h.c.}\,.
\end{eqnarray}
$f_2^c=d a_1^c-ia_1^c\wedge a_1^c$ is the field strength of the $SU(N)$ color field $a_1^c$. $\psi$ and $\tilde \psi$ are two left-handed Weyl fermions in representations ${\cal R}$ and $\overline{\cal R}$ under $SU(N)$, respectively, constituting together a single Dirac fermion. The covariant derivative is $D_\mu=\partial_\mu-i a_\mu^c$ and $y$ is the Yukawa coupling. The potential of the complex scalar field is $V(\Phi)=\lambda\left(|\Phi|^2-v^2/2\right)$, where $\lambda$ is ${\cal O}(1)$ parameter. We take $v\gg \Lambda$, where $\Lambda$ is the strong scale of the gauge sector. The Lagrangian (\ref{YMA lag}) is invariant under two classical $0$-form symmetries $U(1)_B^{(0)}\times U(1)^{(0)}_\chi $, the baryon-number and axial symmetries. The ABJ anomaly in the color background breaks $U(1)_\chi^{(0)}$ down to $\mathbb Z^{\chi(0)}_{2T_{\cal R}}$, and we find that the full good global symmetry of the $SU(N)$ axion-YM theory is \cite{Anber:2019nze,Anber:2020gig}
\begin{eqnarray}\label{Global sym}
G^{\scriptsize \mbox{global}}=\frac{U(1)^{(0)}_B \times \mathbb Z^{\chi(0)}_{2T_{\cal R}}} {\mathbb Z_{N/m}\times \mathbb Z_2^F}\times \mathbb Z_m^{(1)}\,.
\end{eqnarray}
The $1$-form global symmetry $\mathbb Z_m^{(1)}$ acts on Wilson's lines of $a_1^c$, where
 $m=\mbox{gcd}(N,n)$ and $n$ is the $N$-ality of the representation ${\cal R}$, i.e., the boxes in the Young tableaux modulo $N$.  The baryon-number  $U(1)_B^{(0)}$ and the chiral $\mathbb Z^{\chi(0)}_{2T_{\cal R}}$ symmetries act on the local fields as
\begin{eqnarray}\nonumber
U(1)_B^{(0)}&:&\quad \psi\rightarrow e^{i\alpha}\psi\,, \quad \tilde \psi\rightarrow e^{-i\alpha}\tilde\psi\,,\quad \Phi\rightarrow \Phi\,,\\
\mathbb Z^{\chi(0)}_{2T_{\cal R}}&:& \quad \psi\rightarrow e^{i\frac{2\pi \ell}{2T_{\cal R}}}\psi\,, \quad  \tilde\psi\rightarrow e^{i\frac{2\pi \ell}{2T_{\cal R}}}\tilde\psi\,, \quad \Phi\rightarrow e^{-i\frac{4\pi \ell}{2T_{\cal R}}}\Phi\,,
\end{eqnarray}
and $\ell=1,2,..,2T_{\cal R}$ and $T_{\cal R}$ is the Dynkin index of ${\cal R}$ (in our normalization, $T_{\Box}=1$, where $\Box$ is the fundamental representation). The modding by $\mathbb Z_{N/m}\times \mathbb Z_2^F$ in (\ref{Global sym}) is important to remove redundancies. Here, $\mathbb Z_2^F$ is the $(-1)^F$ fermion number subgroup of the Lorentz group\footnote{Notice that the gauge group that faithfully acts on the fermions is $SU(N)/\mathbb Z_m$. Thus, the fermions are charged under $\mathbb Z_{N/m}$ subgroup of the center of $SU(N)$ gauge group. When $N/m$ is even, the fermion number is a subgroup of $\mathbb Z_{N/m}$, i.e., the fermion number is gauged. In this case, all gauge-invariant operators are bosons.}. The complex scalar field can be written as $\Phi=|\Phi|e^{ia}$, where $a$ is the axion. At energy scales $\ll v$, we can set $|\Phi|=v/\sqrt{2}$, and thus, one may only work with the axion as the lightest degree of freedom. 

%%%%%%%%%%%%%%%%%%%%%%%%%%%%%%%%
\subsection{The mixed 't Hooft anomaly and IR Lagrangian}
%%%%%%%%%%%%%%%%%%%%%%%%%%%%%%%%

%%%%%%%%%%%%%%%%%%%%%%
\underline{\bf{ Energy scale $E\gg v$}}
%%%%%%%%%%%%%%%%%%%%%%

Among the anomalies of the axion-YM theory, the mixed anomaly between the $\mathbb Z_m^{(1)}$ $1$-form center and $\mathbb Z^{\chi(0)}_{2T_{\cal R}}$ chiral symmetries is essential in connection with the $3$-form gauge theory. To see the link, we first review this anomaly from the UV point of view \cite{Kapustin:2014gua,Gaiotto:2017yup,Anber:2020xfk,Anber:2021lzb}. We shall be general and examine the anomaly between a subgroup of the full center $\mathbb Z_m^{(1)}$ and chiral symmetries. We shall also work in the Euclidean space. 

To this end,   we turn on a background of $\mathbb Z_p^{(1)}\subseteq \mathbb Z_m^{(1)}$.  This can be implemented by introducing the pair of $U(1)$ fields $(B_1^c,B_2^c)$ and the constraint $p B_2^c=dB_1^c$. Demanding the quantization condition $\int_{{\cal M}_2} d B_1^c\in 2\pi \mathbb Z$ implies the fractional quantization of $B_2^c$ flux:  $\int_{{\cal M}_2}  B_2^c\in \frac{2\pi \mathbb Z}{p}$. We couple $B_2^c$ to fermions as follows. First, we enlarge the gauge group from $SU(N)$ to $U(N)$; we introduce the $\hat a_1^c$ gauge field of $U(N)$ such that $\hat a_1^c\equiv a_1^c+\frac{B_1^c}{p}$ with field strength $\hat f_2^c=d\hat a_1^c+\hat a_1^c\wedge \hat a_1^c$. This, in turn, implies the relation  $\mbox{tr}(\hat f_2^c)=NB_2^c$. Enlarging the group form $SU(N)$ to $U(N)$ introduces an extra degree of freedom, which can be eliminated by postulating the invariance of the theory under the action of an additional $U(1)^{(1)}$ $1$-form gauge symmetry: $\hat a^c_1\rightarrow \hat a^c_1-\lambda_1^c$. This implies that $\hat f_2^c$, $B_1^c$, and $B_2^c$ transform as $\hat f_2^c\rightarrow \hat f_2^c-d\lambda^c_1$, $B_1^c\rightarrow B_1^c- p\lambda_1^c$, and $B_2^c\rightarrow B_2^c -d\lambda_1^c$, such that the condition $dB_1^c=p B_2^c$ remains invariant. 

The mixed anomaly between the $\mathbb Z_p^{(1)}$ center and  $\mathbb Z^{\chi(0)}_{2T_{\cal R}}$ chiral symmetries is envisaged by examining the partition function in the background of the $U(N)$ and $\mathbb Z_p^{(1)}$ fluxes. In such backgrounds, the topological charge is determined by replacing $f_2^c$ with the combination $\hat f_2^c-B^c_2$ in the expression for the topological charge. Importantly, this expression remains invariant under gauge transformations by $\lambda_1^c$. Thus, the topological charge is
\begin{eqnarray}\label{top charge qc}\nonumber
Q^c&=&\frac{1}{8\pi^2}\int_{{\cal M}_4}\mbox{tr}_\Box\left[\left(\hat f_2^c-B^c_2\right)\wedge\left(\hat f_2^c-B^c_2\right)\right]\\
&=&\frac{1}{8\pi^2}\int_{{\cal M}_4}\mbox{tr}_\Box\left[\hat f_2^c\wedge \hat f_2^c\right]-\frac{N}{8\pi^2}\int_{{\cal M}_4}B_2^c\wedge B_2^c\,,
\end{eqnarray}
and is fractional. 
Recalling that $\int_{{\cal M}_4}\mbox{tr}_\Box\left[\hat f_2^c\wedge \hat f_2^c\right]\in 8\pi^2 \mathbb Z$,  the partition function transforms by the phase $-\frac{N}{8\pi^2}\int_{{\cal M}_4}  B_2^c\wedge B_2^c=-\frac{N}{8\pi^2 p^2}\int_{{\cal M}_4} d B_1^c\wedge d B_1^c\in\frac{N\mathbb Z}{p^2}$. This is the mixed anomaly between the $\mathbb Z_p^{(1)}$ $1$-form center and $\mathbb Z_{2T_{\cal R}}^{(0)}$ discrete chiral symmetries. The anomaly is nontrivial, provided that $p^2$ is not a divisor of $N$. It is important to highlight the group-theoretical result 
\begin{eqnarray}\label{GTresult}
\mathbb{Z}_{\scriptsize m/\mbox{gcd}(m,m')}\subseteq\mathbb{Z}_{T_{\cal R}}\,,
\end{eqnarray}
 where we have expressed $N = mm'$. This result can be verified numerically; we shall use it in our analysis below.

%%%%%%%%%%%%%%%%%%%%%%%%%%
\underline{\bf{ Energy scale $\Lambda\ll E\ll v$}}
%%%%%%%%%%%%%%%%%%%%%%%%%%

At energy scale $\Lambda\ll E\ll v$, the magnitude of $\Phi$ freezes and we may set $\Phi\cong \frac{v}{\sqrt 2}e^{ia}$. Also, the fermions acquire a mass $\sim y v$ and decouple. Then, the effective Lagrangian is:
\begin{eqnarray}\label{intermed lag}
{\cal L}_{\Lambda\ll E\ll v}=-\frac{1}{2g^2}\mbox{tr}\left(f_2^c\wedge \star f_2^c\right)+\frac{v^2}{2}da\wedge \star da+T_{{\cal R}}aq^c\,,
\end{eqnarray}
where $q^c$ the topological charge density: $Q^c=\int_{{\cal M}_4}q^c$, and $Q^c$ is given by the expression (\ref{top charge qc}). Thus, we have
\begin{eqnarray}\label{qc charge density}
q^c=\frac{1}{8\pi^2}\left[\mbox{tr}_\Box\left(\hat f_2^c\wedge \hat f_2^c\right)-NB_2^c\wedge B_2^c \right]\,.
\end{eqnarray}
 In particular, one can easily see that the Euclidean version of (\ref{intermed lag}) reproduces the anomaly $e^{-i\frac{2\pi N}{p^2}}$ under the transformation $a\rightarrow a+\frac{2\pi}{T_{\cal R}}$. 
 
In the absence of the center background, the Lagrangian (\ref{intermed lag}) is invariant under the global symmetry group\footnote{In fact, $U(1)^{(2)}$ is only approximate global symmetry. See our discussion after Eq. (\ref{LEEP}).} 
\begin{eqnarray}
G^{\scriptsize \mbox{global}}=\mathbb Z_{T_{\cal R}}^{(0)}\times\left( \mathbb Z_N^{(1)}\tilde{\times} U(1)^{(2)}\right)\,.
\end{eqnarray} 
The $2$-form symmetry $U(1)^{(2)}$ is an emergent winding-number symmetry that acts on axion strings, while $\mathbb Z_{N}^{(1)}$ is an enhanced $1$-form symmetry (remember that the UV genuine $1$-form symmetry is $\mathbb Z_m^{(1)}$) resulting from the decoupling of fermions. Notice that there can be a higher group structure between $\mathbb Z_N^{(1)}$ and $U(1)^{(2)}$ symmetries, and we used the symbol $\tilde{\times}$ to denote this structure. To see it, we activate a background for $\mathbb Z_N^{(1)}$ by introducing the pair $(B_1^{(N)},B_2^{(N)})$ such that $NB^{(N)}_2=dB_1^{(N)}$ and demanding $\int_{M_2}dB_1^{(N)}\in 2\pi\mathbb Z$. This, in turn, implies the flux of $B_2^{(N)}$ is fractional: $\int_{M_2}B_2^{(N)}\in \frac{2\pi \mathbb Z}{N}$.  The pair of fields  $(B_1^{(N)},B_2^{(N)})$ transforms under a $U(1)^{(1)}$ gauge transformation as $B_1^{(N)}\rightarrow B_1^{(N)}+N\lambda_1^{(N)}$ and $B_2^{(N)}\rightarrow B_2^{(N)}+d\lambda_1^{(N)}$, which leaves the relation $NB^{(N)}_2=dB_1^{(N)}$ invariant. We also introduce $C_3$, the background gauge field of the global $U(1)^{(2)}$ symmetry.

 Inspection of (\ref{intermed lag}, \ref{qc charge density}) reveals that the backgrounds of $ \mathbb Z_N^{(1)}$ and $U(1)^{(2)}$ couple to the axion via the term \cite{Choi:2023pdp}
\begin{eqnarray}\label{global coupling a C3}
{\cal L}\supset\frac{1}{2\pi}a G_4\,,
\end{eqnarray}
 where $G_4$ is the field strength of the combined backgrounds. It is given by
 \begin{eqnarray}\label{def G4}
 G_4=dC_3-\frac{T_{\cal R}N}{4\pi}B_2^{(N)}\wedge B_2^{(N)}\,.
 \end{eqnarray}
  $G_4$ is invariant under a gauge transformation by $\lambda_1^{(N)}$ provided that $C_3$ transforms as
  \begin{eqnarray}
  C_3\rightarrow C_3+d\lambda_2+\frac{T_{\cal R}N}{2\pi}\lambda_1^{(N)}\wedge B_2^{(N)}+\frac{T_{\cal R}N}{4\pi}\lambda_1^{(N)}\wedge d\lambda_1^{(N)}\,.
  \end{eqnarray}
The interplay among $C_3$, $B_2^{(N)}$, and $\lambda_1^{(N)}$ indicates a higher-group structure, where $\mathbb Z_N^{(1)}$ represents the daughter symmetry and $U(1)^{(2)}$ the parent symmetry. Notably, the former cannot exist independently of the latter \cite{Cordova:2018cvg}, imposing constraints on the emergent (enhanced) symmetry scales: $E_{\mathbb Z_N^{(1)}}\lesssim E_{U(1)^{(2)}}$. This condition aligns well with effective field theory expectations: $E_{\mathbb Z_N^{(1)}}\cong \sqrt{\lambda}v$, $E_{\mathbb Z_N^{(1)}}\cong yv$, and $\lambda \ll y^2\ll1$; see  \cite{Brennan:2020ehu} for details. 

In a higher-group structure, one cannot gauge the daughter symmetry without gauging the parent. But the reverse is possible. This observation shall play an important role below.  Notice that the higher-group symmetry becomes split (trivialized) if one can write $G_4$ as a total derivative \cite{Choi:2023pdp}. For example, there is no higher-group structure between the genuine $\mathbb Z_m^{(1)}$ symmetry of the UV theory and $U(1)^{(2)}$. To see that, we replace $B_2^{(N)}\wedge B_2^{(N)}$ in Eq. (\ref{def G4}) by $B_2^{c}\wedge B_2^{c}$, where we use the pair $(B_1^c,B_2^{c})$ (which satisfies the constraint $mB_2^c=dB_1^c$) to activate the background of $\mathbb Z_m^{(1)}$.  Thus, $G_4=dC_3-\frac{T_{\cal R}N}{4\pi m^2}dB_1^{c}\wedge d B_1^{c}$. Since $\mathbb Z_{\scriptsize m/\mbox{gcd}(m,m')}\subseteq\mathbb Z_{T_{\cal R}}$ (remember that $N=mm'$), we can write $\frac{T_{\cal R}N}{ m^2}= m''$, $m''\in \mathbb N$, and hence, $G_4=dC_3'\equiv d\left(C_3-\frac{m''}{4\pi}B_1^{c}\wedge d B_1^{c}\right)$, trivializing the higher-group. This observation is important for our subsequent analysis.

%%%%%%%%%%%%%%%%%%%%%%%%
\underline{\bf{ Energy scale $E\ll \Lambda$}}
%%%%%%%%%%%%%%%%%%%%%%%%

Next, we flow to the deep IR at energy scale $\ll \Lambda$, where we assume the theory confines, and hence, the color gauge field is gapped. We must write down an effective Lagrangian that captures the UV center-chiral anomaly. This can be achieved by (i) gauging the $U(1)^{(2)}$ symmetry and introducing the dynamical $3$-form gauge field $c_3$ and (ii) replacing $\mbox{tr}_\Box\left(\hat f_2^c\wedge \hat f_2^c\right)/4\pi$ in Eqs. (\ref{intermed lag}, \ref{qc charge density}) by $dc_3$:
\begin{eqnarray}\label{low lag}\nonumber
{\cal L}_{E\ll\Lambda}=\frac{v^2}{2}da\wedge \star da+\frac{T_{{\cal R}}a}{2\pi}\left(dc_3-\frac{N}{4\pi}B_2^c\wedge B_2^c\right)+\Lambda^4{\cal K}\left(\frac{dc_3-\frac{N}{4\pi}B_2^c\wedge B_2^c}{\Lambda^4}\right)\,,\\
\end{eqnarray}
and we added a kinetic energy term for $c_3$. The field strength of $c_3$ satisfies the quantization condition $\int_{{\cal M}_4} dc_3\in 2\pi\mathbb Z$, which is simply the infrared manifestation of the quantization of topological charges in Yang-Mills theory.  The reader will notice that the coupling between $a$ and $dc_3$ has an extra factor of $T_{\cal R}$ compared to the coupling in Eq. (\ref{global coupling a C3}). This is because  $c_3$ in Eq. (\ref{low lag}) is a dynamical rather than a background field, and as the dynamical field absorbs the axion, it should describe the formation of $T_{\cal R}$ domain walls. As we shall discuss later, at an energy scale below $\Lambda$, the theory has enhanced $\mathbb Z_N^{(1)}$ $1$-form symmetry. As noted above, we are allowed to gauge $U(1)^{(2)}$ without gauging the daughter symmetry $\mathbb Z_N^{(1)}$. This is important; otherwise, we would have changed the theory's global structure and run into trouble since $\mathbb Z_N^{(1)}$ is not a genuine symmetry of the theory.

  The Lagrangian (\ref{low lag}) must pass several checks. First, it must be invariant under the $\mathbb Z_{2T_{\cal R}}^{\chi (0)}$ chiral symmetry in the absence of the center background, which is evident from the transformation $a \rightarrow a + \frac{2\pi}{T_{\cal R}}$ along with the condition $\int_{{\cal M}_4} dc_3\in 2\pi\mathbb Z$. Second, the Lagrangian must be invariant under the same auxiliary $U(1)^{(1)}$ gauge transformation, by $\lambda_1^c$, of the UV theory. This is the case provided that $c_3$ transforms as 
\begin{eqnarray}\label{transformation of c3}
c_3\rightarrow c_3+d\lambda_2 +\frac{p'}{2\pi }B_1^c\wedge d\lambda_1^c+\frac{pp'}{4\pi}\lambda_1^c\wedge d\lambda_1^c\,,
\end{eqnarray}
and we wrote $N=pp'$. Also, the Lagrangian (\ref{low lag}) must reproduce the mixed $\mathbb Z_{2T_{\cal R}}^{\chi(0)}$-$\mathbb Z_p^{(1)}$ anomaly of the UV theory. This can be easily verified by observing that the partition function acquires the phase $e^{-i\frac{2\pi N}{p^2}}$ when $a$ is shifted by $a \rightarrow a + \frac{2\pi}{T_{\cal R}}$ in the presence of the center background. In the absence of a center background, the Lagrangian (\ref{low lag}) exactly matches (\ref{full axion lag}) in Section \ref{Gauging the 2-form symmetry: Domain walls}, and everything we said there applies here.

Another check\footnote{We thank T. Tanizaki for pointing this out.} on the validity of (\ref{low lag}) is to integrate out $c_3$ along the lines of our discussion that led from Eq. (\ref{path int over c3}) to Eq. (\ref{alt der}).  Thus, we sum over arbitrary values of the integers $\int_{{\cal M}_4} f_4\in 2\pi\mathbb Z$ and use the Poisson resummation formula $\sum_{m\in \mathbb Z}\delta\left(2\pi m-\int_{{\cal M}_4} f_4\right)=\sum_{k\in \mathbb Z}e^{-ik\int_{{\cal M}_4}f_4}$. We also use the change of variables $\hat f_4=f_4-\frac{N}{4\pi}B_2^c\wedge B_2^c$. Specifying to the canonical kinetic term ${\cal K}_{\scriptsize \mbox{can}}$,  performing the Gaussian integral over $\hat f_4$, and focusing only on the zero modes, we obtain the Euclidean partition function
\begin{eqnarray}\label{eff in back}
Z[a]\sim\sum_{k\in \mathbb Z} e^{-i\frac{k N}{4\pi}\int_{{\cal M}_4}B_2^c\wedge B_2^c}e^{-\frac{\Lambda^4 V_{{\cal M}_4}}{8\pi^2}\left(T_{\cal R}a+2\pi k\right)^2}\,.
\end{eqnarray}
This effective partition function picks up the anomaly $e^{i\frac{ N}{4\pi}\int_{{\cal M}_4}B_2^c\wedge B_2^c}=e^{-i\frac{2\pi N}{p^2}}$ upon shifting $a\rightarrow a+\frac{2\pi}{T_{\cal R}}$. It also displays the expected structure of the Yang-Mills theory: it has an infinite number of vacua, with the true vacuum energy density given by 
\begin{eqnarray}\label{LEEP}
V(a)=\frac{\Lambda^4}{8\pi^2}\mbox{min}_k\left(T_{\cal R}a+2\pi k\right)^2\,.
\end{eqnarray}
The potential $V(a)$ has $T_{\cal R}$ minima with cusps at $a=\pi/T_{\cal R}, 3\pi/T_{\cal R}$, etc.. The cusps indicate that the potential $V(a)$ is missing degrees of freedom at these locations. These are the hadronic walls sandwiched between the axion domain walls. An axion wall has width $\sim v/\Lambda^2$, while a hadronic wall is much thinner with width $\sim \Lambda^{-1}$. Including the infinite sum over all the integers $\int_{{\cal M}_4} f_4\in 2\pi m$, $m \in \mathbb Z$ was crucial to see these cusps. As emphasized above, the integer $m$ is the IR manifestation of the Yang-Mills instantons' topological charge. Below $\Lambda$, the theory is strongly coupled, and the vacuum receives contributions from all the topological charge sectors. 

Let us examine the theory's behavior at energy scales $\Lambda \ll E \ll v$, such as at a corresponding temperature. In this case, it suffices to include the contribution from minimal charges $\int_{{\cal M}_4} f_4 = m = \pm 1$ in the partition function. Translating this into the language of Yang-Mills instantons, the dilute instanton-gas approximation is reliable at this temperature because it serves as an infrared cut-off on the instanton's scale modulus \cite{Gross:1980br}. Thus, summing over the smallest instantons, which possess topological charges of $\pm 1$, is adequate. Limiting the sum over $m$ to the lowest charge sector means that one can no longer perform the Poisson resummation that leads to (\ref{eff in back}), and thus, one can no longer make sense of (\ref{LEEP}) or the cusps. This is consistent with the expectation that the hadronic walls melt away at a temperature $\sim \Lambda$. Nevertheless, at temperatures in the range $\Lambda \ll T \ll v$, the classical theory (\ref{low lag}) still possesses $T_{\cal R}$ vacua with axion domain walls interpolating between them. Notice that in this energy range, $c_3$ does not strongly fluctuate (since$\int_{{\cal M}_4} f_4 =\pm 1$), and we can consider the $U(1)^{(2)}$ $3$-form gauge field $c_3$ as a background rather than a dynamical field. Thus, one may still regard $U(1)^{(2)}$ as an approximate global symmetry. Eventually, the axion domain walls melt at temperature $T \gtrsim v$. 

As noted in  \cite{Luscher:1978rn,Gabadadze:2002ff,Dvali:2005an} and discussed in the Introduction, it was recognized that the IR behavior of both pure YM theory in the large-$N$ limit and the axion-YM theory can be effectively described using the $3$-form gauge field $c_3$. Here, incorporating $c_3$ in our discussion has been essential for aligning with the infrared constraints of the 't Hooft anomaly. Our method provides a systematic approach to argue for a consistent infrared effective field theory of the axion-YM system.

At energy scales below $\Lambda$, the theory acquires the global symmetry
 \begin{eqnarray}
 G^{\scriptsize \mbox{global}}=\mathbb Z_{T_{\cal R}}^{(0)}\times\left( \mathbb Z_N^{(1)}\tilde{\times} \mathbb Z_{T_{\cal R}}^{(3)}\right)\,.
 \end{eqnarray} 
  In general, a higher-group structure may exist between $\mathbb Z_N^{(1)}$ and $\mathbb Z_{T_{\cal R}}^{(3)}$, which becomes apparent when both symmetries' backgrounds are activated. The background of $\mathbb Z_N^{(1)}$ was discussed earlier, while that of $\mathbb Z_{T_{\cal R}}^{(3)}$ can be activated by introducing the pair $(F_3, F_4)$, satisfying the constraint $dF_3=T_{\cal R}F_4$, along with the quantization condition $\int_{{\cal M}4} dF_3\in 2\pi\mathbb Z$ \cite{Tanizaki:2019rbk}. The axion coupled to these backgrounds is represented by
\begin{eqnarray}
{\cal L}\supset \frac{a}{2\pi}\left(T_{\cal R}dc_3+dF_3-\frac{ T_{\cal R}N}{4\pi }B_2^{(N)}\wedge B_2^{(N)}\right).
\end{eqnarray}
Maintaining invariance under a gauge transformation by $\lambda_1^{(N)}$ requires $F_3$ to transform as
\begin{eqnarray}
F_3\rightarrow F_3+d\lambda_2+\frac{T_{\cal R}N}{2\pi}d\lambda_1^{(N)}\wedge B_2^{(N)}+\frac{T_{\cal R}N}{4\pi}d\lambda_1^{(N)}\wedge d\lambda_1^{(N)}\,.
\end{eqnarray}
The interplay among $F_3$, $B_2^{(N)}$, and $\lambda_1^{(N)}$ indicates a higher-group symmetry $\mathbb Z_N^{(1)}\tilde{\times} \mathbb Z_{T_{\cal R}}^{(3)}$. However, this higher-group structure becomes trivial if $dF_3-\frac{N T_{\cal R}}{4\pi N^2 }dB_1^{(N)}\wedge dB_1^{(N)}$ can be expressed as a total derivative. This holds particularly true for the $\mathbb Z_m^{(1)}$ symmetry, as demonstrated above in a similar case involving $U(1)^{(2)}$ and $\mathbb Z_m^{(1)}$. Understanding this aspect is pivotal when gauging the genuine center, as this operation should be executed without gauging $\mathbb Z_{T_{\cal R}}^{(3)}$.

In the IR, the symmetries  $\mathbb Z_{T_{\cal R}}^{(0)}$ and $\mathbb Z_{T_{\cal R}}^{(3)}$ undergo spontaneous breaking. The enhanced symmetry $\mathbb Z_N^{(1)}$ remains unbroken until length scales $\sim yv/\Lambda^2$, at which point it also undergoes explicit breaking due to the heavy fermions that pop up from vacuum as we take the Wilson lines to be larger than $\sim yv/\Lambda^2$. This leaves $\mathbb Z_m^{(1)}$ as the sole surviving unbroken symmetry.

%%%%%%%%%%%%%%%%%%%%%%%%%%%%%%%%%%%
\subsection{$SU(N)/\mathbb Z_p$ and noninvertible chiral symmetry}
%%%%%%%%%%%%%%%%%%%%%%%%%%%%%%%%%%%%%

Let us investigate whether our construction yields the desired results when we gauge the genuine center or any of its subgroups, aligning with the well-established findings in the literature \cite{Kaidi:2021xfk,Argurio:2023lwl,Anber:2023mlc}. We shall see that the answer is affirmative, lending support to the picture that the deep IR regime of the system is genuinely described by the $3$-form gauge theory. 

We consider the same axion-YM theory with matter, but now let us gauge a subgroup of the center $\mathbb Z_p^{(1)}\subseteq\mathbb Z_m^{(1)}$, i.e., we consider $SU(N)/\mathbb Z_p$ axion-YM theory with matter\footnote{In principle, there are $p$ distinct theories: $(SU(N)/\mathbb Z_p)_k$, $k=0,1,..,p-1$ differing by the admissible genuine (electric, magnetic, or dyonic) line
operators \cite{Aharony:2013hda}. In this paper, we limit our treatment to $(SU(N)/\mathbb Z_p)_{k=0}$. The Hilbert space and the noninvertible chiral symmetry in $(SU(N)/\mathbb Z_p)_{k=0}$ theory were considered in \cite{Anber:2023mlc}.}. This theory is constructed by promoting $( B_1^c, B_2^c)$ to dynamical fields $(b_1^c, b_2^c)$ and performing the sum over the fractional instantons in the path integral. Let us define the new $3$-form gauge field $\hat c_3$:
\begin{eqnarray}
\hat c_3\equiv c_3-\frac{N}{4\pi p^2} b_1^c\wedge d b_1^c\,,
\end{eqnarray}
keeping in mind the quantization condition $\int_{{\cal M}_4} dc_3\in 2\pi\mathbb Z$.
The Lagrangian of this theory at energy scale $E\ll \Lambda$ reads
\begin{eqnarray}\label{full IR axion L gauging p}
{\cal L}_{E\ll \Lambda}[(b_1^c,b_2^c)]=\frac{v^2}{2}da\wedge \star da+\frac{T_{\cal R}}{2\pi}a\wedge d\hat c_3 +\Lambda^4{\cal K}\left(\frac{d\hat c_3}{\Lambda^4}\right)\,,
\end{eqnarray}
and the partition function is
\begin{eqnarray}
Z=\sum_{(b_1^c,b_2^c)}\int [Dc_3][Da]e^{i\int_{{\cal M}_4} {\cal L}_{E\ll \Lambda}[(b_1^c,b_2^c)]}\,.
\end{eqnarray}
In the Kalb-Ramond frame, we replace $q\rightarrow T_{\cal R}$ and $c_3\rightarrow \hat c_3$ in (\ref{KR manin Lag}). It is important to repeat what we stated above: we can gauge the genuine $\mathbb Z_m^{(1)}$ center symmetry or a subgroup thereof without spoiling $\mathbb Z_{T_{\cal R}}^{(3)}$ since the pair does not constitute a higher-group. On the contrary, gauging the enhanced $\mathbb Z_N^{(1)}$ is disastrous: it entails that we also gauge $\mathbb Z_{T_{\cal R}}^{(3)}$, which destroys the domain walls.

The chiral symmetry defect  is given by (\ref{sym of 0-form}) after replacing $c_3$ with $\hat c_3$ and summing over $(b_1^c, b_2^c)$:
\begin{eqnarray}
 \tilde U_{\ell}^{(0)}(\mani_{3})\sim\sum_{( b_1^c,b_2^c)} e^{i \frac{2\pi \ell}{T_{\cal R}} \int_{\mani_{3}} v^{2} \star da-i \ell \int_{{\cal M}_3} \left( c_3-\frac{N}{4\pi p^2} b_1^c\wedge d b_1^c\right) }\,,\quad  \ell =1,2,.., T_{{\cal R}}\,.
\end{eqnarray}
The new defect $ \tilde U_\ell^{(0)}(\mani_3)$ defines a noninvertible chiral symmetry $\tilde{\mathbb Z}_{T_{\cal R}}^{(0)}$. To simplify the form of $ \tilde U_{\ell}^{(0)}(\mani_3)$, we write $N$ as $N=pp'$ and assume that $p'=1$ Mod $p$. Then, $\frac{N}{4\pi p^2} b_1^c\wedge d b_1^c\sim \frac{1}{4\pi p} b_1^c\wedge db_1^c$, i.e., this is an improperly quantized quantum Hall term. We may rewrite it in terms of an auxiliary $1$-form gauge field $\varphi_1$ that lives on ${\cal M}_3$:
\begin{eqnarray}
 \tilde U_{\ell}^{(0)}(\mani_{3})\sim   \sum_{(b_1^c, b_2^c)} e^{i \frac{2\pi \ell}{T_{\cal R}} \int_{\mani_{3}} v^{2} \star da-i\ell \int_{{\cal M}_3}c_3}\int [D\varphi_1] e^{-i\ell \int_{{\cal M}_3} \left(\frac{p}{4\pi}\varphi_1\wedge d\varphi_1+\frac{1}{2\pi} \varphi_1\wedge d b_1^c\right) }\,.
\end{eqnarray}
The last term is the minimal abelian TQFT ${\cal A}^{p,1}$ discussed in \cite{Hsin:2018vcg}. When $\ell$ is a multiple of $p$, the TQFT is trivialized, giving us an invertible symmetry.

%%%%%%%%%%%%%%
\section{Discussion}
\label{Discussion}
%%%%%%%%%%%%%%

In this letter, we critically evaluated the hypothesis that the $3$-form gauge theory offers more than an alternative framework for the deep IR regime of axion-Yang-Mills systems. We commenced by rigorously investigating the $3$-form gauge theory coupled to axions, as encapsulated by the Lagrangian (\ref{full axion lag}). Our analysis revealed that this theory represents a network of domain walls terminating on an axion string, with particular emphasis placed on its global structure. A dual formulation, the Kalb-Ramond Lagrangian (\ref{KR manin Lag}), was also considered, which describes the same physical phenomenon. Notably, while some symmetry defects are explicit in one formulation, others are evident in the alternative frame. Crucially, in the absence of gravitational effects\footnote{Inclusion of gravity may differentiate between the axion and Kalb-Ramond frames, see \cite{Duff:1980qv}.}, both formulations are equivalent, possessing identical global symmetries.

Subsequently, we examined the $SU(N)$ Yang-Mills theory with a Dirac fermion coupled to an $SU(N)$-neutral complex scalar, highlighting the necessity of an emergent $3$-form gauge field for IR matching of the theory's mixed center-chiral anomaly.  Consider varying the complex scalar vev such that we go to the limit $v\ll \Lambda$. In this case, the strong dynamics set in well before the axions are amenable to the weak-coupling treatment. In this opposite limit, the theory still forms domain walls leading to $T_{\cal R}$ distinct vacua, and thus, we do not expect a bulk phase transition to take place as we vary $v$ below or above $\Lambda$. We may not rigorously justify the introduction of the $3$-form gauge field in this scenario. However, by continuity, we expect that the $3$-form gauge theory remains a valid description in the deep IR. This reasoning, combined with the large-$N$ limit analysis discussed in the introduction, supports the notion that the vacuum of Yang-Mills theory should likely be described by a $3$-form gauge theory.

Incorporating gravity offers further insights into the significance of the $3$-form description, as the cosmological constant in this context can be interpreted as arising from a gauge principle. Theoretically, one can distinguish between a pure cosmological constant and a $3$-form gauge theory, as the latter yields a non-vanishing contribution to the trace anomaly proportional to the Gauss-Bonnet invariant \cite{Duff:1980qv,Christensen:1978md}. 
Irrespective of this subtle effect, Brown and Teitelboim realized that the action (\ref{Luscher action}) taken as a starting point with no reference to its UV completion leads to the quantum creation of closed membranes localized on the boundary of ${\cal M}_4$ \cite{Brown:1987dd,Brown:1988kg}. As the membranes are produced, the vacuum energy density associated with $c_3$ decreases, reducing the effective value of the cosmological constant. This idea, when refined, may lead to a solution to the cosmological constant problem \cite{Bousso:2000xa,Bousso:2007gp}. Also, connections between the QCD vacuum and the cosmological constant problem were discussed in \cite{Kaloper:2017fsa,Urban:2009vy}. 

As discussed in the paper, introducing the $3$-form gauge field $c_3$ was necessary to match the chiral-center anomaly in an axion-Yang-Mills system. Eventually, $c_3$ eats the axion, becoming a short-range field, and the cosmological constant vanishes. Yet, one can think of an alternative scenario with an axion, two distinct Yang-Mills fields, and a chiral-center anomaly. In this case, two $3$-form gauge fields are anticipated. One combination of these fields eats the axion, while an orthogonal combination remains gapless. The latter can source a cosmological constant. Intriguingly, in this scenario, the infrared cosmological constant can be considered a by-product of the 't Hooft anomaly-matching condition. However, in such a scenario, global symmetries should only be considered approximate since exact global symmetries are forbidden in quantum gravity, see, e.g., \cite{Misner:1957mt,Banks:2010zn,Harlow:2018tng}. 

The method presented in this paper can be extended in various ways. One immediate application is to address the problem of multi-flavor quarks by incorporating the $3$-form gauge theory description in the chiral Lagrangian. Another venue is applying our formalism to the Standard Model (SM) and its variants, potentially through coupling with axions. It is well-established that the SM exhibits a $\mathbb{Z}_6^{(1)}$ $1$-form symmetry, and the true gauge group might be modded by $\mathbb{Z}_6$ or a subgroup thereof (see, e.g., \cite{Tong:2017oea,Anber:2021upc,Wang:2018jkc,Wan:2019gqr,Li:2024nuo,Alonso:2024pmq,Koren:2024xof}). Exploring whether and how a $3$-form gauge theory may emerge deep in the IR of the SM and its extensions and whether the modded discrete group could play a significant role in this formalism will be an exciting avenue of research. Additionally, linking the emergent $3$-form to the observed cosmological constant presents another intriguing possibility. These investigations are worthy of future exploration.

%%%%%%%%%%%%%%%%%%%%%%%%%%%%%%%%%%%%%%%%%%%%%%%%%%%%%%%%%%%%%%%%%
{\bf {\flushleft{Acknowledgments:}}}  We thank M. Bullimore, A. Grigoletto, N. Iqbal, T. Jacobson, E. Poppitz, T. Sulejmanpasic, Y. Tanizaki,  M. \"{U}nsal, and A. Zhitnitsky for many discussions. We also thank E. Poppitz for his comments on the Manuscript. We would also like to thank an anonymous referee for their valuable suggestions, which helped improve the discussion.   This work is supported by STFC through grant ST/T000708/1. 

{\em This work is dedicated to the memory of my late father, Mostafa, whose love, support, and encouragement have been a constant source of inspiration.} M.A.
%%%%%%%%%%%%%%%%%%%%%%%%%%%%%%%%%%%%%%%%%%%%%%%%%%%%%%%%%%%%%%%%%%

  \bibliography{RefKRGlobal.bib}
 
  \bibliographystyle{JHEP}

  \end{document}